\documentclass[aps,prd,10pt,nofootinbib,twocolumn,eqsecnum,showpacs,showkeys,superscriptaddress,preprintnumbers,altaffilletter]{revtex4-1}

\usepackage[T1]{fontenc}
\usepackage[utf8]{inputenc}
\usepackage{graphicx}
\usepackage{amsmath}
\usepackage{amssymb}
\usepackage{amsfonts}
\usepackage{amsbsy}
\usepackage{multirow}
\usepackage{booktabs}
\usepackage{soul}

\usepackage{hyperref}
\hypersetup{
    colorlinks=true,
    linkcolor=blue,
    filecolor=magenta,
    urlcolor=cyan,
    citecolor=cyan
}

\newcommand{\be}{\begin{equation}}
\newcommand{\ee}{\end{equation}}
\newcommand{\bea}{\begin{eqnarray}}

\begin{document}

\title{DHOST gravity in Ultra-diffuse galaxies - Part II: NGC 1052-DF4 and Dragonfly 44}

\author{E. Laudato}
\email{enrico.laudato@phd.usz.edu.pl}
\affiliation{Institute of Physics, University of Szczecin, Wielkopolska 15, 70-451 Szczecin, Poland}
\author{V. Salzano}
\email{vincenzo.salzano@usz.edu.pl}
\affiliation{Institute of Physics, University of Szczecin, Wielkopolska 15, 70-451 Szczecin, Poland}

\date{\today}

\begin{abstract}

Ultra-Diffuse galaxies are a family of gravitational systems with quite varied properties. On one hand we have cases like NGC1052-DF2 and NGC1052-DF4, both observed by the Dragonfly Array Telescope, which are claimed to be highly-deficient in dark matter. On the other hand, we have also observed Ultra-Diffuse galaxies which are almost totally dominated by dark matter, such as DF44, which is estimated to be at $99\%$ dark. To explain such kind of a variety of behaviors might be a problem for both the standard dark matter paradigm and for alternative theories of gravity that try to overcome dark matter existence by modifying General Relativity. 

Here we consider a modified gravity theory belonging to the family of Degenerate Higher-Order Scalar Tensor theories to study the internal kinematics of both NGC1052-DF4 and DF44. The peculiarity of the chosen model is the partial breaking of its Vaishtein screening mechanism for which it might have an influence not only on cosmological scales but also on astrophysical ones. We consider two different possibilities: one in which the model only plays the role of dark energy; and another one in which it might also mimic a sort of effective dark matter. 

We get conflicting results. For NGC1052-DF4 we confirm that the galaxy dynamics might be successfully described even only by a stellar component and that, at least at the scale which are probed, the content of dark matter is quite low. In addition to that, we also show that the alternative gravity model is totally consistent with data and is statistically equivalent to a standard General Relativity dark matter scenario, and it might even replace dark matter. On the contrary, DF44 requires dark matter both in General Relativity and in our alternative gravity model. When the latter is considered only as a cosmological dark energy fluid, it is statistically fully reliable and equivalent to General Relativity. But when we try to use it to substitute dark matter, although we get good fits to the data, the constraints on the theoretical parameters are in sharp contrast with those derived from more stringent probes from the stellar scales.

\end{abstract}

\maketitle

\section{Introduction}

Ultra-Diffuse galaxies (UDGs) are galaxies with an extremely low surface brightness. Although recently re-named after a dedicated survey performed with the Dragonfly Array telescope \citep{vanDokkum:2014cea,2014PASP..126...55A} centered on the Coma Cluster, they have been known since a longer time and are not a new galaxy type, but may be considered as a specific subset of dwarf spheroidals and dwarf ellipticals \cite{2018RNAAS...2...43C}. 

They are characterized by half-light radii $R_{eff} = 1.5-4.6$ kpc and central surface brightness $\mu_0 = 24-26$ mag arcsec$^{-2}$ (in the $g$ photometric band); their sizes resemble spiral galaxies such as the Milky Way; their luminosities appear like those of elliptical galaxies \citep{2003MNRAS.343..978S}; however, their stellar populations are more similar to dwarfs galaxies. UDGs can be localized in dense environments \citep{vanDokkum:2014cea,2015ApJ...813L..15M}, in galaxy groups \citep{2017MNRAS.468.4039R,2017ApJ...842..133L} and also in voids \citep{2019MNRAS.486..823R,2019MNRAS.485.1036M}. 

Their presence in different environments implies many different formation scenarios. In \citep{10.1093/mnras/stz383,10.1093/mnras/staa854} it is reported that they could have been formed due to tidal stripping processes. In \citep{10.1093/mnrasl/slw055,10.1093/mnras/stx1440} it is suggested they could be from dark matter (DM) halos characterized by high angular momentum. Since UDGs are poor-gas galaxies \citep{vanDokkum:2014cea}, they might be galaxies that have lost their gas component after forming the first generation of stars (as a consequence of supernovae, AGN feedback  \citep{2015ApJ...804...18A,2013ApJ...775..116R} and gas stripping \citep{10.1093/pasj/56.1.29,2015MNRAS.452..937Y}), or after ``galaxy harassment'' \citep{2014PASP..126...55A}.

Furthermore, UDGs are usually characterized by a high number of globular clusters (GCs). This abundance appears unusual for such faint galaxies \citep{2016ApJ...830...23B,2016ApJ...822L..31P} and points towards massive DM halos \citep{2016ApJ...830...23B,2016ApJ...822L..31P,Harris:2013zea}).

DF44, one of the largest UDGs observed with the Dragonfly Array telescope, represents the best case for such a scenario. The galaxy is characterized by a large population of GCs $\sim 100$ \citep{vanDokkum:2016uwg} (although the real number might be smaller, as shown in \citep{vanDokkum:2019fdc}) which implies an halo mass $sim 10^{11} M_\odot$ or, equivalently, an estimated $\sim99\%$ of the total mass.

However, the case of DF44 cannot be generalized to all UDGs. Indeed, in \citep{vanDokkum:2018vup} the UDG NGC1052-DF2 (DF2) has been pointed to be a ``lacking DM'' galaxy, compatible with the hypothesis of a purely baryonic galaxy. In \citep{2019ApJ...874L...5V}, a second galaxy, NGC1052-DF4 (DF4), has been reported to have a very low amount of DM. The lack of DM of DF2 and DF4 could represent, as stated in \citep{vanDokkum:2018vup}, a falsification of the standard cosmological scenario or might raise questions about the galaxy formation processes that require DM as the essential component \citep{Silk:1967kq}. DF2 has been studied in many aspects, but the debate is still open (see \cite{Laudato:2022vmq} for a summary). 


Here we want to approach the problem of the co-existence of UDGs like DF44, DF4, and DF2 from the point of view of modified gravity theories. The dynamics of DF44 has been studied in \citep{vanDokkum:2019fdc, Wasserman:2019ttq} assuming General Relativity (GR) but ``fuzzy'' dark matter \citep{Duffy:2009ig} as an alternative to the standard candidate, weakly interacting massive particles (WIMPS) \citep{Bertone:2004pz,Queiroz:2017kxt}. Both DF44 and DF2 have been analysed in \citep{Islam:2019irh,Islam:2019szu} within three different gravity scenarios, namely, the Weyl conformal gravity \citep{Mannheim:1988dj,Mannheim:2005bfa}, the MOND (with and without assuming the External Field Effect \citep{Hees:2015bna}) and Moffat's Modified Gravity (MOG) \cite{Moffat:2005si}. It emerges that while the Weyl and the MOND (with the External Field Effect) scenarios could explain the velocity dispersion data of DF2 and DF44 without invoking DM, the MOG fails in describing the kinematics of DF44. These results stress that gravity modifications might represent a viable path to describe and explain UDGs' structure in a unified picture.

In this work, we aim to continue what we started in \citep{Laudato:2022vmq} for DF2, by considering both DF4 and DF44, in the context of a particular Degenerate Higher-Order Scalar Tensor (DHOST) theory \citep{PhysRevD.97.021301,Dima:2017pwp,BenAchour:2016fzp,Langlois:2018dxi,Crisostomi:2016czh} to test its reliability to explain the dynamics of gravitational structures at different scales. DHOST theories are mainly intended to reproduce dark energy (DE) on large scales, and are characterized by a screening mechanism \citep{Vainshtein:1972sx,Brax_2013,Koyama:2015vza,Burrage:2017qrf} needed to recover GR at smaller Solar System scales \citep{PhysRevLett.92.121101,Bertotti:2003rm}. What is peculiar of many DHOST models is that such screening may be spontaneously broken \citep{Kobayashi:2014ida}; therefore, the gravity modification introduced to play the role of DE might turn, on astrophysical scales, into the effects commonly attributed to DM. 

In \citep{10.1093/mnras/stac180} we have shown that  such DHOST is compatible with data from clusters of galaxies; in \citep{Laudato:2022vmq}, we moved to galactic scales and considered the case of DF2 \citep{2018Natur.555..629V}. 
Here we will study a similar (to DF2) case, DF4 (here for the first time), and a totally opposite object, DF44, adopting two scenarios: \textit{a)} the DHOST model as an alternative to DE only; \textit{b)} the DHOST model representing an alternative to also DM due to the partial breaking of the Vainshtein screening. Testing opposite systems like DF2 and DF4 on one end, and DF44 on the other one, will allow us to understand if it could be reliable mimicking DM
distribution in galaxies with a modified gravity scenario in which DM can be seen just as an effective gravitational effect induced by some specific property of the chosen modified gravity (in our case, the partially broken screening mechanism).

The paper is organized as follows: in Sec.~\ref{sec: Model} we introduce all the technicalities about the DHOST model and the internal kinematics arguments we will use for the analysis of DF4 and DF44; in Sec.~\ref{sec:modelling} we explain how we model the mass components of the two galaxies; in Sec.~\ref{sec: statistics} we illustrate how we perform the statistical analysis; in Sec.~\ref{sec: results} we present the results achieved underling the implications for what concern the two scenarios studied within the DHOST framework; in Sec.~\ref{sec: conclusions} we finally derive the conclusion of this work.

In this work we assume that the DHOST model behaves on cosmological scales as a standard $\Lambda$CDM model  with $H_0 =67.74$ km s$^{-1}$Mpc$^{-1}$, $\Omega_m =0.3089$ and $\Omega_\Lambda = 0.6911$. Our choice for a $\Lambda$CDM background finds a qualitative support in \citep{Creminelli:2017sry} and a more quantitative one in \citep{Hiramatsu:2020fcd,Hiramatsu:2022fgn}.

\section{Theory}
\label{sec: Model}

The model belonging to the family of Degenerate Higher-Order Scalar Tensor Theories (DHOST), introduced in \cite{Langlois:2015cwa,Crisostomi:2016czh, PhysRevD.97.021301, PhysRevD.97.101302} is characterized by a partial breaking of the Vainshtein screening mechanism \cite{Vainshtein:1972sx} whose net effect is the modification of the gravitational ($\Phi$) and the metric potentials ($\Psi$) as
\begin{align}
\label{eqn: model}
\frac{d\Phi}{dr} &= \frac{G_NM(r)}{r^2} + \Xi_1 G_N M''(r)\, , \nonumber \\
\frac{d\Psi}{dr} &= \frac{G_NM(r)}{r^2} + \Xi_2 \frac{G_N M'(r)}{r} + \Xi_3 G_N M''(r)\, ,
\end{align}
with: $G_N$, the measured effective gravitational constant (the bare constant $G$ defined in terms of the Planck mass $M_{Pl} = (8\pi G)^{-1}$); $M(r)$, the spherical mass enclosed in the radius $r$ and $M'(r)$ and $M''(r)$, respectively, the first and the second order derivatives of the mass with respect to $r$; $\Xi_{1,2,3}$, the parameters that define the properties of the model \cite{PhysRevD.103.064065}. From Eq.~\eqref{eqn: model} it is possible to see that GR is recovered for $\Xi_{1,2,3}\to 0$. The most general definitions of the three parameters $\Xi_{1,2,3}$ can be given within the Effective Field Theory approach \cite{Dima:2017pwp}. Given the recent constraints coming from the multi-messenger gravitational waves observation of GW170817 \citep{Creminelli:2017sry, LIGOScientific:2017ync}), their expressions can be substantially simplified \citep{Creminelli:2017sry,Sakstein:2017xjx,Ezquiaga:2017ekz,Baker:2017hug,Baker:2020apq} as:
\begin{eqnarray}
\label{eqn: xi param}
\Xi_1 &=& -\frac{1}{2}\frac{(\alpha_H+\beta_1)^2}{\alpha_H+2\beta_1}\, ,\\
\Xi_2 &=& \alpha_H\, ,\\
\Xi_3 &=& -\frac{\beta_1}{2}\frac{(\alpha_H+\beta_1)}{\alpha_H+2\beta_1}\, , \\
\gamma_0 &=& - \alpha_H - 3 \beta_1\, ,
\end{eqnarray}
where $\gamma_0$ represents the fractional difference between $G_N$ and $G$; $\alpha_H$ measures the kinetic mixing between matter and the scalar field \citep{Gleyzes:2014qga}; and $\beta_1$ accounts for higher-order terms in the Lagrangian \citep{Langlois:2018dxi}. We recover $\Lambda$CDM limits with $\alpha_H, \beta_1\to 0$. 

\subsection{Internal Kinematics of UDGs}

In order to study the kinematics of our UDGs, we start from the Jeans equation under the assumption of dynamical equilibrium and spherical symmetry:
\be
\label{eqn: Jeans}
\frac{d(l\sigma_r^2)}{dr} + \frac{\beta(r)}{r}(l\sigma_r^2) = l(r)\frac{d\Phi(r)}{dr}\, ,
\ee
where $l(r)$ represents the luminosity density of the galaxy (expressed in $L_\odot$ kpc$^{-3}$), $\beta(r)$ is the anisotropy parameter \citep{2008gady.book.....B}, and $\Phi$ is the gravitational potential, whatever the gravity theory is. Therefore, Eq.~\eqref{eqn: Jeans} allows us to study the compatibility of our chosen modified gravity scenario with observational data.

Within GR we can solve Eq.~\eqref{eqn: Jeans} by integration, thus getting the radial velocity dispersion $\sigma_r$ \cite{Mamon:2004xk}:
\be
\label{eqn: solution}
l(r)\sigma^2_r(r) = \frac{1}{f(r)}\int^\infty_rds\hspace{.1em}f(s)l(s)\frac{M(s)}{s^2}\, ,
\ee
where the function $f(r)$ depends on the functional form of the anisotropy parameter through the definition
$d\log f(r)/d\log r = 2\beta(r)$. The observed quantity, although, is the line-of-sight projection of $\sigma_r$, i.e.
\begin{align}
\label{eqn: vlos}
\sigma^2_{los}(R) = &\frac{2}{I(R)}\biggl[\int^\infty_Rdr\hspace{.1em}r\frac{l\sigma^2_r}{\sqrt{r^2 - R^2}} - \nonumber\\
&R^2\int^\infty_Rdr\hspace{.1em}\beta(r)\frac{l\sigma^2_r}{r\sqrt{r^2 - R^2}}\biggr]\, ,
\end{align}
where $R$ is the 2D projected radius, and $I(R)$ is the surface density of the galaxy. A more compact  definition of Eq.~\eqref{eqn: vlos} has been presented in \citep{Mamon:2004xk} as
\be
\label{eqn: finvlos}
\sigma^2_{los}(R) = \frac{2G_N}{I(R)}\int^\infty_Rdr\hspace{.1em}K\biggl(\frac{r}{R}\biggr)l(r)\frac{M(r)}{r}\, ,
\ee
where the Kernel function $K(r/R)$ can be fully specified only after the assumption of a specific parametrization of the anisotropy parameter \citep{Mamon:2004xk,Mamon:2012yb}.

At this point, Eq.~\eqref{eqn: finvlos} must be conveniently modified in order to be used with the chosen DHOST model. For that goal, it is possible to see that Eq.~\eqref{eqn: model} can be rewritten as
\be
\frac{d\Phi}{dr} = \frac{G_NM_{eff}(r)}{r^2}\, ,
\ee
thus introducing a sort of effective mass $M_{eff} = M(r) + \Xi_1 r^2 M''(r)$ to take into account the new effects of the DHOST with respect to GR. By using this simple redefinition of the mass we can rewrite Eq.~\eqref{eqn: finvlos} within the DHOST framework, as
\be
\label{eqn: finvlos_DHOST}
\sigma^2_{los}(R) = \frac{2G_N}{I(R)}\int^\infty_Rdr\hspace{.1em}K\biggl(\frac{r}{R}\biggr)l(r)\frac{M_{eff}(r)}{r}\, .
\ee

\section{UDGs mass modeling}
\label{sec:modelling}

The UDG galaxies DF4 \citep{2019ApJ...874L...5V} and DF44 \citep{vanDokkum:2016uwg,vanDokkum:2019fdc} belong to the 47 low surface brightness galaxies observed with the Dragonfly telescope \citep{vanDokkum:2014cea} in the Coma cluster.

\subsection{Stellar component}

For both galaxies DF4 and DF44, we model the stellar component as a single S\'{e}rsic profile \citep{1963BAAA....6...41S}
\be
\label{eqn: lum}
I(R) = I_{0} \exp \left[ -\left( \frac{R}{a_{s}}\right)^{1/n}\right]
\ee
where: $I_0$ is the central surface brightness; $a_s$ is the S\'{e}rsic scale parameter (expressed in kpc); $n$ is the S\'{e}rsic index. We can relate the S\'{e}rsic scale parameter $a_s$ with the half-to-light radius $R_{eff}$ according to
\be
a_s = \frac{R_{eff}}{(b_n)^n}
\ee
where $b_n$ is a function of the S\'{e}rsic index $n$ defined as $b_n = 2n -0.33$ \citep{Caon:1993wb}. For the parameters which describe the stellar component we consider:
\begin{itemize}
    \item DF4: from \citep{2019ApJ...874L...5V} we have: S\'{e}rsic index $n = 0.79$; effective radius $R_{eff} \sim 16.58$ arcsec, derived from the given value of $1.6$ kpc at a fiducial distance $D = 20$ Mpc; axis ratio $b/a = 0.89$; central surface brightness $\mu_0 = 23.7$ mag arcsec$^{-2}$. All photometric quantities are in the $V_{606}$ band. At the given fiducial distance the absolute magnitude $M_{V}= -15.0$ converts into a total luminosity of $L_{tot} = 7.7\cdot 10^7 L_\odot$;
    \item DF44: from \citep{vanDokkum:2019fdc, Wasserman:2019ttq} we have: S\'{e}rsic index $n = 0.94$; effective radius $R_{eff} \sim 9.69$ arcsec, calculated from the value of $4.7$ kpc at the fiducial distance $D = 100$ Mpc; axis ratio $b/a = 0.68$; central surface brightness $\mu_0 = 24.1$ mag arcsec$^{-2}$. Once again, all photometric quantities are in the $V_{606}$ band. At the distance given fiducial distance the absolute magnitude $M_{V}= -16.2$ corresponds to a total luminosity $L_{tot} = 2.33\cdot 10^8 L_\odot$. 
\end{itemize}
From the surface brightness $I(R)$ it could be possible to obtain the luminosity density profile $l(r)$ by deprojection, but an exact analytical expression for general S\'{e}rsic index $n$ does not exist. In alternative, an analytical approximation is supplied in \citep{1997A&A...321..111P}, namely
\be
\label{eqn: lumin}
l(r) =l_1\widetilde{l}(r/a_s)\, ,
\ee
where $\widetilde{l}(r/a_s)$ is a dimensionless function defined as
\be
\widetilde{l}(x) = x^{-p_n}\exp(-x^{1/n})\, ,
\ee
with $p_n \simeq 1 - 0.6097/n + 0.05463/n^2$ \citep{Neto:1999gx}, and
\be
\label{eqn: elle1}
l_1 = \frac{L_{tot}}{4\pi n \Gamma[(3 - p_n)n]a^3_s}\, .
\ee
The total luminosity is explicitly expressed in terms of the distance as
\be
\label{eq:ltot}
L_{tot} = 10^{-0.4(m_{V_{606}} - \mu(D) - M_{\odot,V_{606}})}
\ee
where: $m_{V_{606}}$ is the apparent magnitude of the galaxy; $\mu(D) = 5\log_{10}D + 25$ is the distance modulus (the distance $D$ of the galaxy is measured in Mpc); $M_{\odot,V_{606}} = 4.72$ is the Sun's absolute magnitude in the $V_{606}$ photometric band \cite{2018ApJS..236...47W}. Finally, the correct dimensionality of the mass density of the stellar component is recovered by multiplying equation Eq.~(\ref{eqn: lumin}) by the mass-to-light ratio $\Upsilon_*$, 
\be
\label{eqn: stardens}
\rho_*(r) = \Upsilon_*l\left(r\right)\, .
\ee
Integrating Eq.~\eqref{eqn: stardens}, within a certain radius $r$, one can get the analytical expression of the total mass for the stellar component as
\begin{align}
\label{eqn: stellar mass}
&M_*(<r) = 2\pi n \Upsilon_* I_0 \left(\frac{R_{eff}}{b^n_n}\right)^2\frac{\Gamma(2n)}{\Gamma[(3 - p_n)n]} \times\\
&\left\{\Gamma[(3 - p_n)n] - \gamma\left[(3 - p_n)n,b_n\left(\frac{r}{R_{eff}}\right)^{1/n}\right]\right\}\, ,
\end{align}
with $\Gamma(z) = \int_0^\infty dt\hspace{.1 em}t^{z - 1}e^{-t}$ and $\gamma(z,x) = \int_x^\infty dt\hspace{.1 em}t^{z - 1}e^{-t}$, respectively, the total and the upper incomplete gamma functions.

The impossibility of effectively measuring the anisotropy profile $\beta(r)$, and the relative degeneracy, can be broken only by assuming some functional form for $\beta(r)$. In this work, we consider two different choices: constant anisotropy, $\beta(r) \equiv \beta_c$; and a radial anisotropy applied to UDGs in \citep{Zhang:2015pca} and defined as
\be
\label{eqn: anis}
\beta(r) = \beta_0 + (\beta_\infty - \beta_0)\frac{r}{r + r_a}\, ,
\ee
where $\beta_0$ and $\beta_\infty$ are the inner and the outer anisotropy for $r\to 0$ and $\infty$ respectively, and $r_a$ is a scale parameter. 


\subsection{Dark matter}

For the DM component we consider a generalized Navarro-Frenk-White (gNFW) profile \citep{Zhao:1995cp,Moore:1999gc}
\be
\label{eqn: gNFW}
\rho_{gNFW}(r) = \rho_s\biggl(\frac{r}{r_s}\biggr)^{-\gamma}\biggl(1 + \frac{r}{r_s}\biggr)^{\gamma - 3}
\ee
where $\rho_s$ and $r_s$ are, respectively, the scale density and the scale length of the density profile \citep{Navarro:1995iw} and $\gamma$ represents the inner log-slope. The classical NFW density profile is recovered for $\gamma = 1$. 
The choice of parametrizing the DM density profile with a gNFW profile is mainly motivated by the fact that in \citep{Wasserman:2018scp,10.1093/mnras/stab3491,Wasserman:2019ttq,Brook_2021} the standard NFW profile seems to be unsuitable for UDGs, and a gNFW profile has more freedom to adjust the data.

The total mass enclosed in a given radius $r$ can be calculated and reads
\begin{align}
M_{DM}(<r) &= \frac{4 \pi \rho_{s} r^{3}_{s}}{3-\gamma} \left(\frac{r}{r_{s}} \right)^{3-\gamma}\, \\ 
&_2F_1\left[3 - \gamma,3- \gamma,4 - \gamma,- \frac{r}{r_s}\right]\, , \nonumber
\end{align}
where $_2F_1$ is the hypergeometric function.

Instead of using $\{\rho_s,r_s\}$ we parametrize the DM density in terms of the concentration parameter $c_{\Delta} = r_{\Delta}/r_s$ and of the virial mass $M_{\Delta} = 4\pi/3\, \Delta\, \rho_{c}r^{3}_{\Delta}$, with $\Delta = 200$, i.e., all the quantities are evaluated at the radius where the density of the system is $200$ times the critical density $\rho_c$ of the Universe,
\be
\rho_c = \frac{3H^2(z)}{8\pi G}\, ,
\ee
where $H(z)$ is the Hubble parameter. For what concerns the redshift, DF4 belongs to the NGC 1052 group, then we consider $z = 0.004963$\footnote{from the NED database:  \url{https://ned.ipac.caltech.edu/}}. For DF44, we consider the redshift of the Coma Cluster $z= 0.023156$ from the SIMBAD catalog \citep{Wenger:2000sw}.
Given $\{M_{200},c_{200}\}$ the scale density $\rho_s$ can be expressed as
\be
\label{eqn: rhos}
\rho_s = \frac{200}{3}\rho_{c}(z) \frac{(3-\gamma)(c_{200})^\gamma}{_2F_1\left[3 - \gamma,3- \gamma,4 - \gamma,- c_{200}\right]}\, .
\ee

\section{Statistical analysis}
\label{sec: statistics}



Seven compact objects similar to globular clusters (GCs) have been observed in DF4  using the dual arm Low-Resolution Imaging Spectrograph (LRIS) on the Keck I telescope. 
The kinematical data reported in Table 1 of \citep{2019ApJ...874L...5V} include GC's radial velocities and the corresponding uncertainties. Thus, the $\chi^2$ for DF4 is defined as
\be
\label{eqn: chiDF4}
\chi^2(\boldsymbol{\theta}) = \sum_{i}^{\mathcal{N}_{data}}  \frac{(v_i - v_{sys})^2}{\sigma_{i}^2} + \ln \left(2\pi \sigma_{i}^2\right)
\ee
where: $\mathcal{N}_{data}=7$ is the total number of the data points; $v_i$ are the observed GCs velocities; $v_{sys}$ is the systemic velocity of the galaxy;
$\sigma_{i}^2 = \sigma^2_{los,i}(\boldsymbol{\theta}) + \sigma^2_{v_i}$ is the total error on the velocities $v_{i}$, with $\sigma_{v_{i}}$ the measurement uncertainties and $\sigma^2_{los,i}(\boldsymbol{\theta})$ the velocity dispersion, which explicitly depends on the model parameters, $\boldsymbol{\theta}$ through Eqs.~(\ref{eqn: finvlos}) and (\ref{eqn: finvlos_DHOST}).

We additionally apply three Gaussian priors: on the distance, $D = 22.1 \pm 1.2$ Mpc, as measured in \citep{Shen:2021zka} using the tip of the red-giant branch (TRGB) method \citep{1993ApJ...417..553L}; on the mass-to-light ratio $\Upsilon_* = 2.0 \pm 0.5\, M_\odot/L_\odot$ \citep{2018Natur.555..629V}; and on the systemic velocity $v_{sys} = 1444.6 \pm 7.75$ km s$^{-1}$ \citep{2019ApJ...874L...5V}. On the anisotropy parameters $\beta_i$ (with $i = \{c,0,\infty\}$) we consider log-normal priors $-\log (1 - \beta_i) = 0. \pm 0.5$ over the range $[-10,1]$. 


DF44's spectroscopy was obtained through the Keck Cosmic Web Imager (KCWI) on the Keck II telescope. It possesses a large population of globular clusters ($\sim$ 100) and a stellar velocity dispersion $\sigma = 47^{+8}_{-6}$ kms$^{-1}$, a larger value than those of other UDGs observed with the Dragonfly telescope, like DF2 and DF4. 
The kinematical data we have used are those from Table 2 of \citep{vanDokkum:2019fdc}, where radial circular velocities $v_{i}$ and dispersion $\sigma_{i}$, with corresponding uncertainties, are given for nine radial bins. In this case, the $\chi^2$ reads
\be
\label{eqn: chiDF44}
\chi^2(\boldsymbol{\theta}) = \sum_{i = 1}^{\mathcal{N}_{data}}\left(\frac{\sigma_{eff,i} - \sigma_{los,i}(\boldsymbol{\theta})}{\delta\sigma_i}\right)^2 + \ln(2\pi\delta\sigma^2_i)
\ee
where: $\mathcal{N}_{data}= 9$ is the number of the data points; $\sigma_{eff,i} = (\sigma^2_i + v^2_i)^{1/2}$ is the effective velocity dispersions  \citep{vanDokkum:2019fdc}; and $\delta\sigma_{eff,i}$ are the corresponding uncertainties. 

Also for DF44 we apply some priors: one Gaussian on the distance, assumed to be that of the Coma Cluster, $D = 102\pm 14$ Mpc, determined in \citep{Thomsen:1997vz} using the Surface Brightness Fluctuation (SBF) method \citep{1988AJ.....96..807T,2010ApJ...724..657B}; one log-normal on the mass-to-light ratio $\log\Upsilon_* = 1.5$  \citep{Wasserman:2019ttq} with a scatter of $0.1$ dex \citep{2011MNRAS.418.1587T}; the same  log-normal priors as DF4 on the anisotropy parameters. Finally, for both galaxies, we put a control to guarantee that $\sigma_{los} > 0$.



According to the actual galaxy formation paradigm, galaxies are embedded in DM halos which are expected to extend far beyond the range of distances provided by data. To address this problem and ensure a physical significance for dark matter parameters, we set a log-normal prior on the concentration parameter $c_{200_c}$ using the $c-M$ relation provided by \citep{Correa:2015dva}, updated at the latest cosmology from \textit{Planck} 2015 cosmology (see their Appendix B1), and covering a range of masses and redshifts which makes it more suitable for our galaxies than other relations in literature. The applied dispersion is $\sigma_{\log c_{200}} = 0.16$ dex.   

We also consider two scenarios for the virialized mass $M_{200}$: one in which the DM halo and the stellar component are related through the Stellar-to-halo-mass relation (SHMR) \citep{2017MNRAS.470..651R} and one in which the two components are decoupled. In the former case, we assume a log-normal prior on $M_{200}$, whose mean is provided by \citep{2017MNRAS.470..651R} and scatter of $0.3$ dex. For the latter one, we assume a uniform flat prior in the range $\log_{10} M_{200}/M_\odot \in [2,15]$.

No priors have been imposed on the DHOST parameter $\Xi_1$, leaving it as a free parameter.

\begin{figure*}
\centering
\includegraphics[width=8.cm]{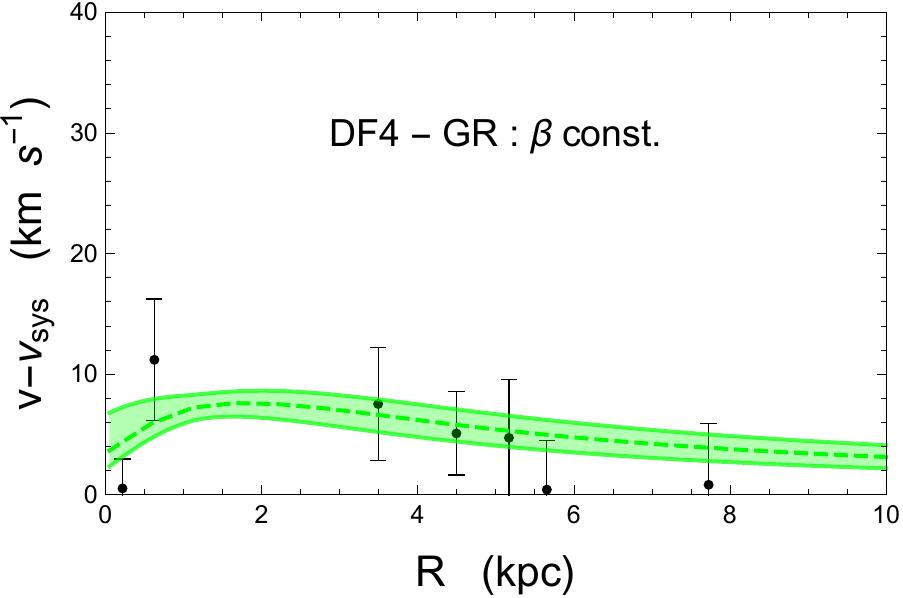}~~~
\includegraphics[width=8.cm]{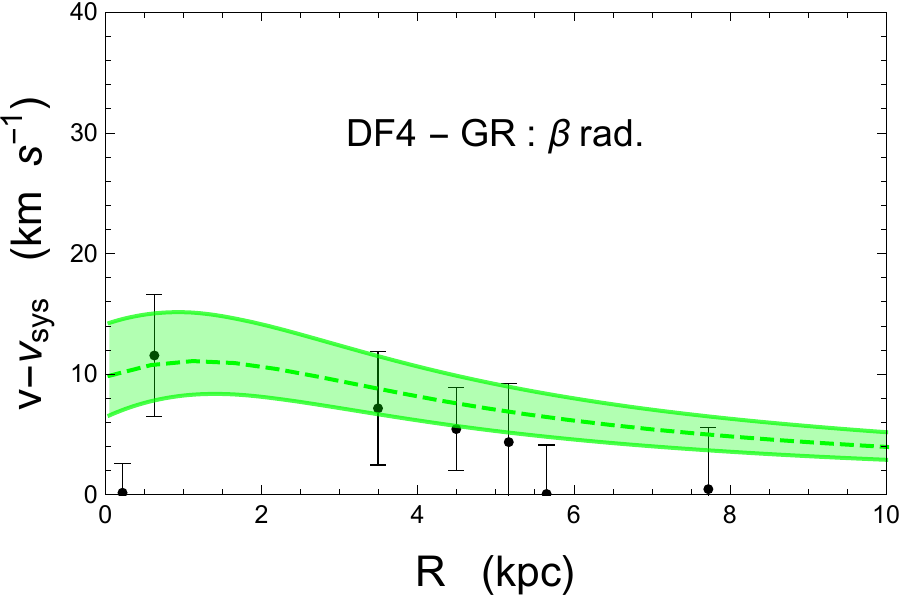}\\
~~~\\
\includegraphics[width=8.cm]{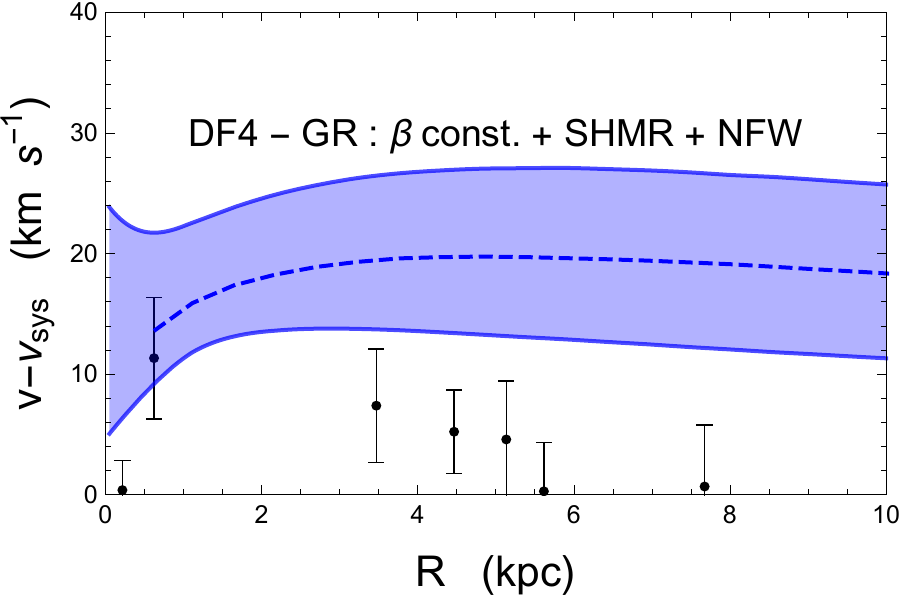}~~~
\includegraphics[width=8.cm]{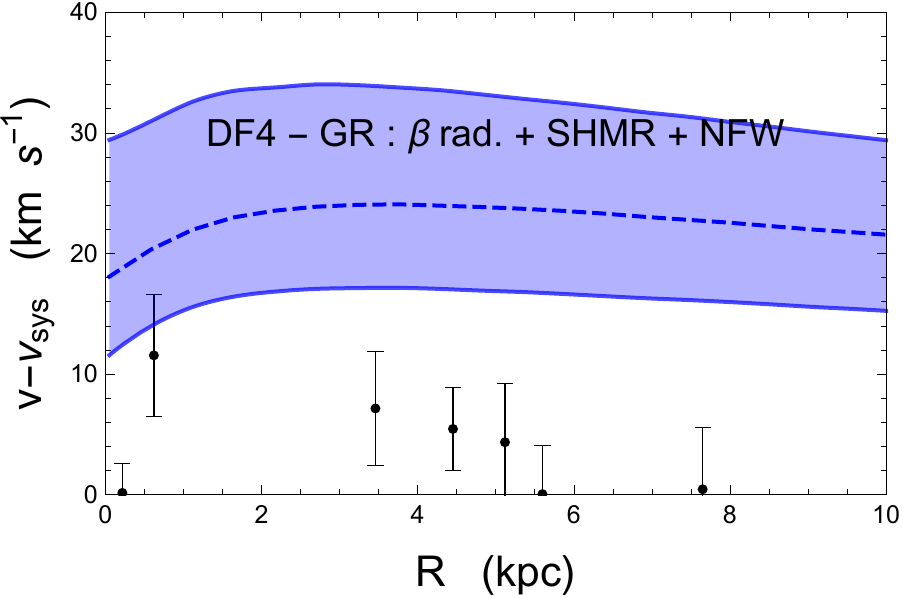}\\
~~~\\
\includegraphics[width=8.cm]{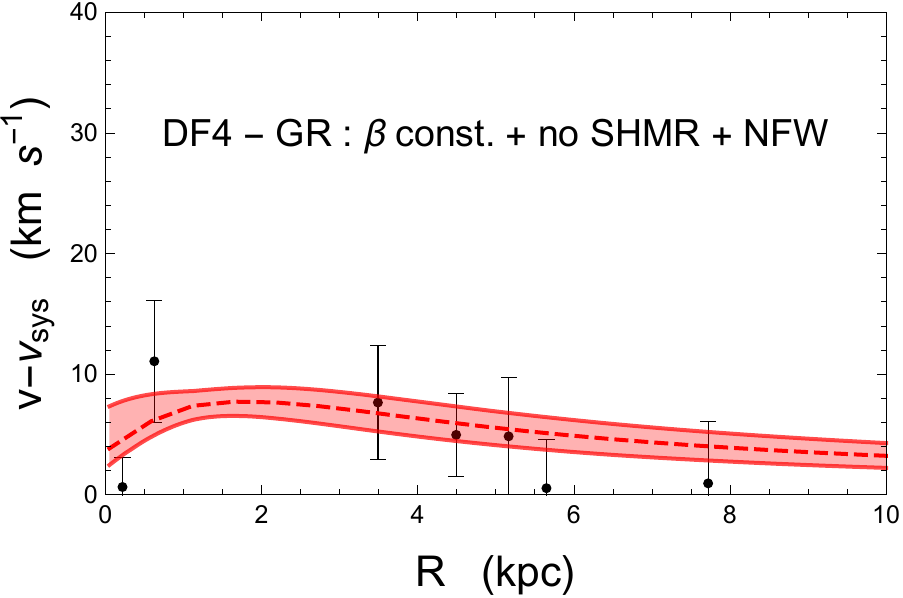}~~~
\includegraphics[width=8.cm]{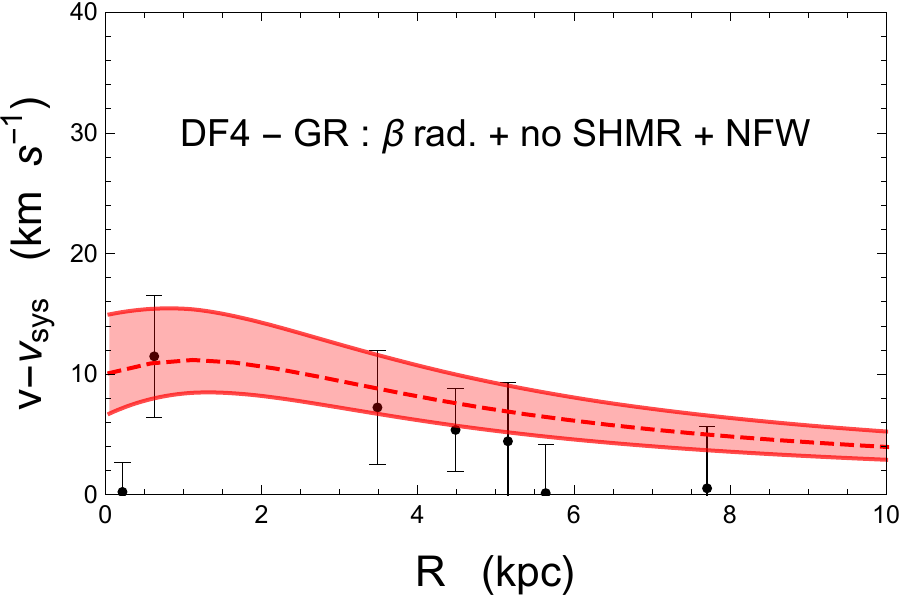}\\
\caption{Velocity offset profiles of NGC1052-DF4 with GR. Black dots and bars are observational data, $v_{i}-v_{sys}$, with uncertainties, $\sigma_{v_{i}}$. Colored dashed lines and shaded regions are respectively the median and the $1\sigma$ confidence region of the $\sigma_{los}$ profile derived from Eq.~\eqref{eqn: finvlos}. Left panel: constant velocity  anisotropy profile. Right panel: radial velocity profile.}\label{fig:plot_DF4_GR}
\end{figure*}


Finally, we minimize the $\chi^2$ using our own code for running Monte Carlo Markov Chain (MCMC). The chain convergence is checked using the method introduced in \citep{Dunkley:2004sv}. In order to establish a statistically reliable hierarchy of scenarios that better agree with the data, we rely on the Bayes Factor \citep{doi:10.1080/01621459.1995.10476572}, $\mathcal{B}^{i}_{j}$, defined as the ratio between the Bayesian Evidences of model/scenario $\mathcal{M}_i$ and model/scenario $\mathcal{M}_j$. The evidence is calculated numerically using our own code implementing the Nested Sampling algorithm developed by \citep{Mukherjee:2005wg}. The Bayes Factor is then interpreted in terms of the empirical Jeffrey's scale \citep{Jeffreys1939-JEFTOP-5}. 

However, the Bayes Factor is a prior-dependent quantity \citep{Nesseris:2012cq}. Indeed, decreasing/increasing the widths of priors can lead to higher (lower) evidence, increasing (decreasing) the tensions between models $\mathcal{M}_i$ and $\mathcal{M}_j$. Although we have made the choice to include only physical priors and, when possible, on the largest range possible, we also determine quantities that do not depend on the priors (or whose dependence is very weak). For that, we use the Suspiciousness $\mathcal{S}^{i}_{j}$ introduced in \citep{Handley:2019wlz,Handley:2019pqx,Joachimi:2021ffv} and defined as
\be
\label{eqn: sus}
\log\mathcal{S}^i_j = \log\mathcal{B}^i_j + \mathcal{D}_{KL,i} - \mathcal{D}_{KL,j}
\ee
where $\log\mathcal{B}^i_j$ is the logarithm of the Bayes Factor; and $\mathcal{D}_{KL}$ is the Kullback-Leibler (KL) divergence \citep{10.1214/aoms/1177729694}. An interpretation of the suspiciousness, similar to Jeffrey's scale for the Bayes Ratio, is provided by Fig. 4 of \citep{Joachimi:2021ffv}. Specifically, a negative value of $\log\mathcal{S}^i_j$ should be intended as a sign of tension; a positive value of $\log\mathcal{S}^i_j$ instead as a sign of concordance.

\section{Results}
\label{sec: results}

All the results of our analysis may be found in Table ~\ref{tab:results1} for DF4 and Table ~\ref{tab:results2} for DF44. As for the interpretation of the Bayes Factor and Suspiciousness, the reference scenario with respect to which we compare all our cases is GR with only baryonic matter for DF4, and GR with gNFW DM profile and a SHMR prior for DF44.

\subsection{NGC 1052-DF4}

We first start analyzing the GR scenario. DF4, as well as DF2, is seemingly characterized by a deficiency of DM within the range of distances probed by observations \citep{2019ApJ...874L...5V}.

\begin{figure*}
\centering
\includegraphics[width=8.cm]{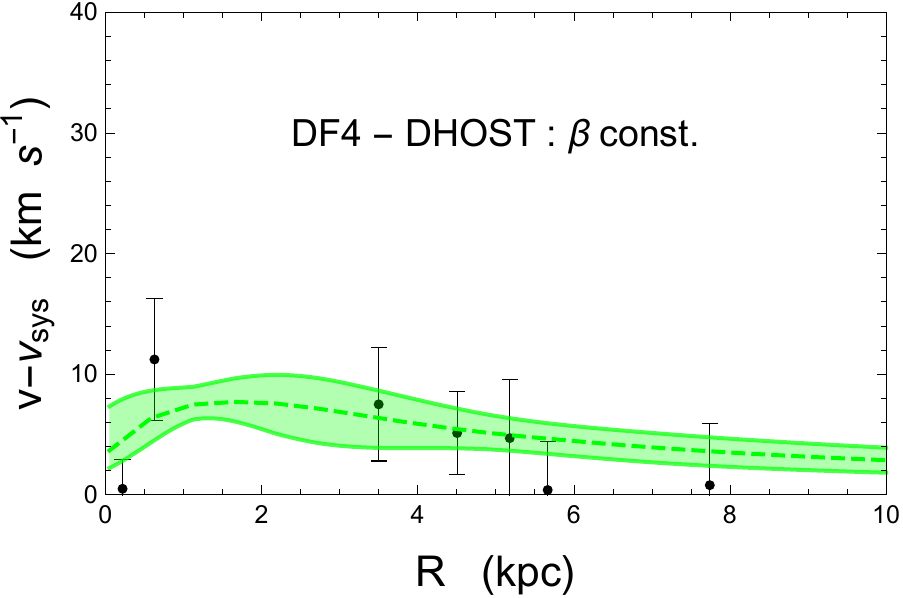}~~~
\includegraphics[width=8.cm]{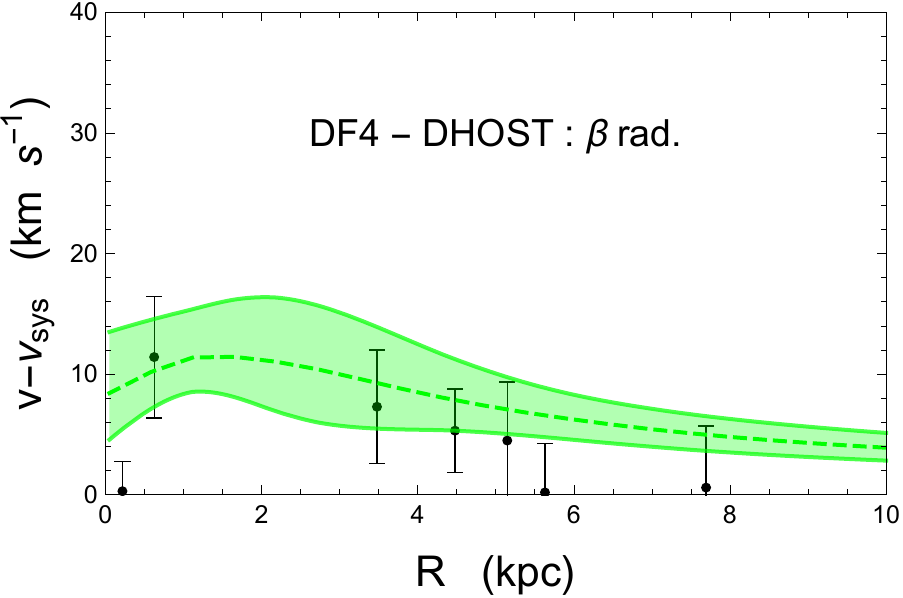}\\
~~~\\
\includegraphics[width=8.cm]{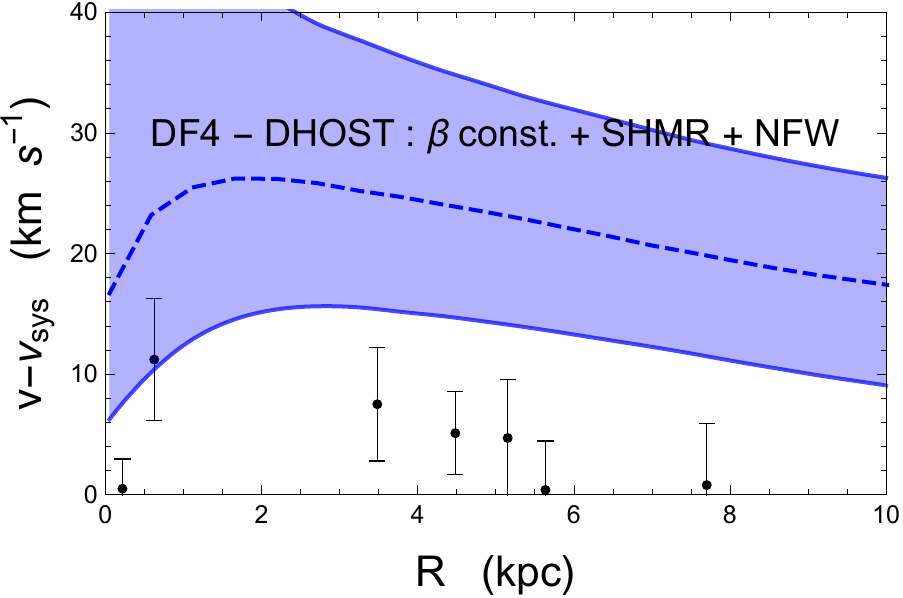}~~~
\includegraphics[width=8.cm]{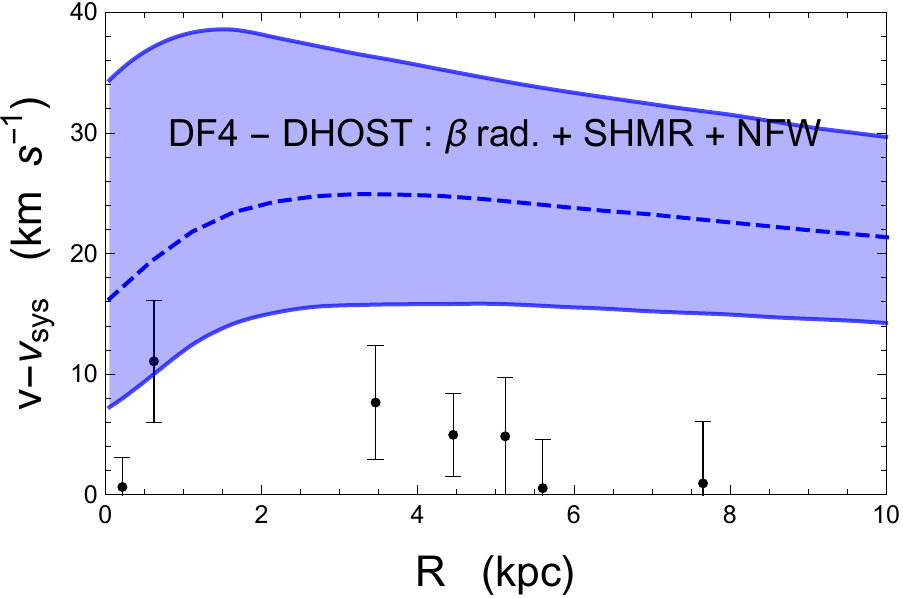}\\
~~~\\
\includegraphics[width=8.cm]{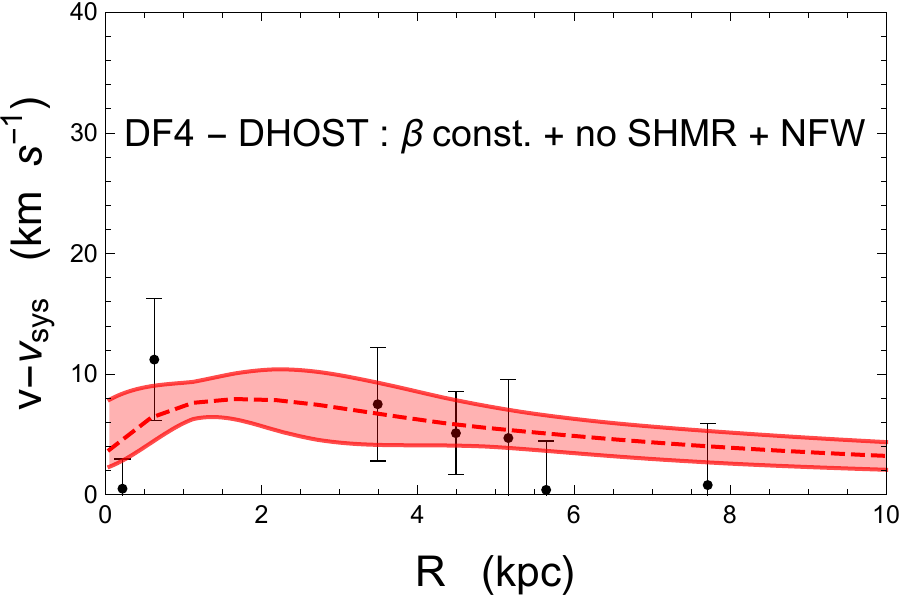}~~~
\includegraphics[width=8.cm]{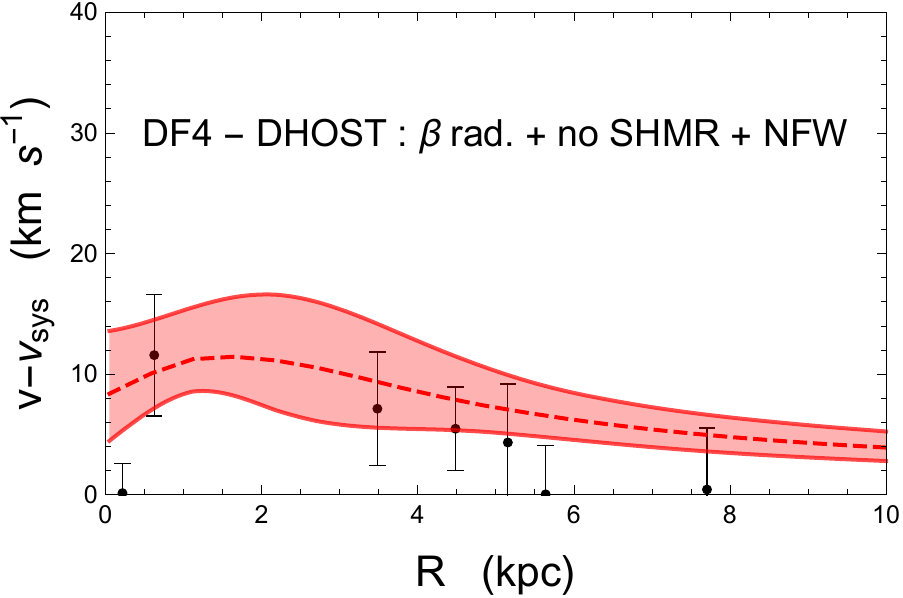}
\caption{Velocity offset profiles of NGC1052-DF4 with DHOST. Black dots and bars are observational data, $v_{i}-v_{sys}$, with uncertainties, $\sigma_{v_{i}}$. Colored dashed lines and shaded regions are respectively the median and the $1\sigma$ confidence region of the $\sigma_{los}$ profile derived from Eq.~\eqref{eqn: finvlos_DHOST}. Left panel: constant velocity anisotropy profile. Right panel: radial velocity profile.}\label{fig:plot_DF4_DHOST}
\end{figure*}

\begin{figure*}
\centering
\includegraphics[width=8.cm]{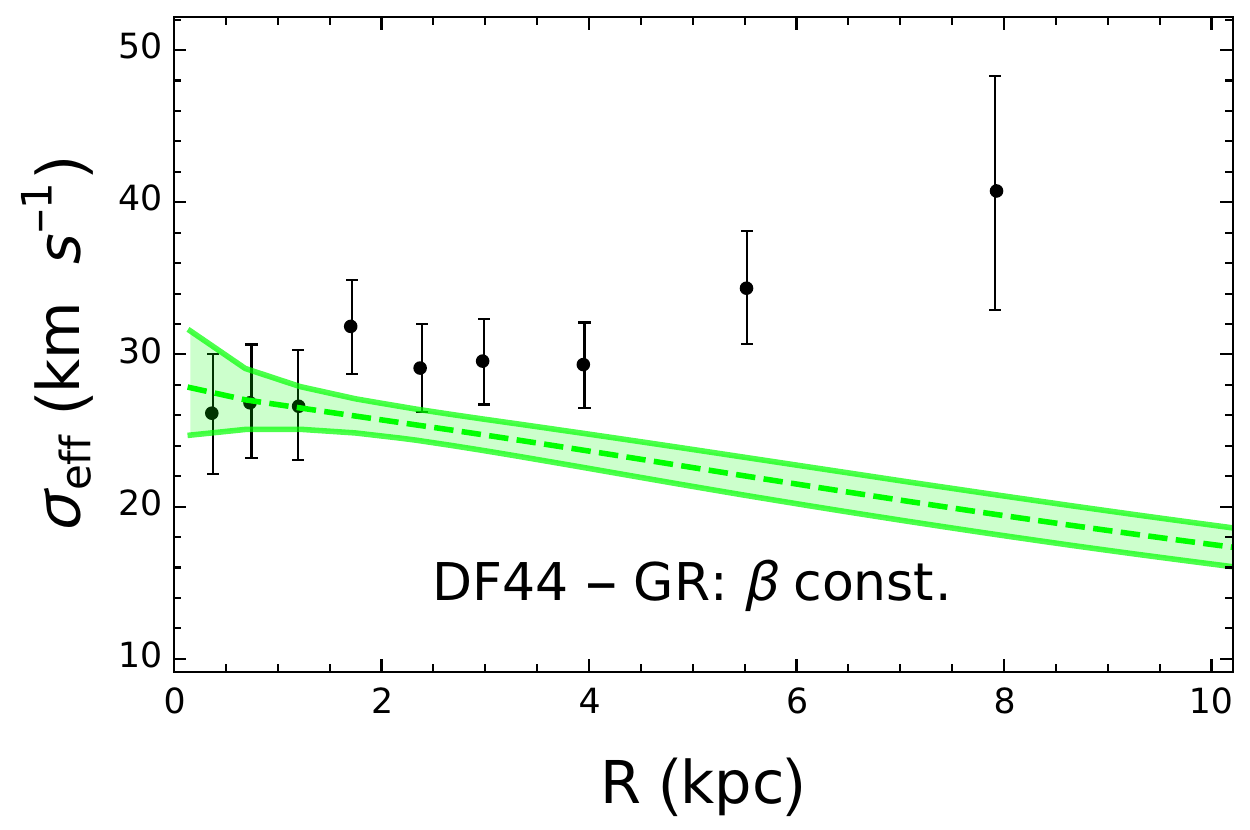}~~~
\includegraphics[width=8.cm]{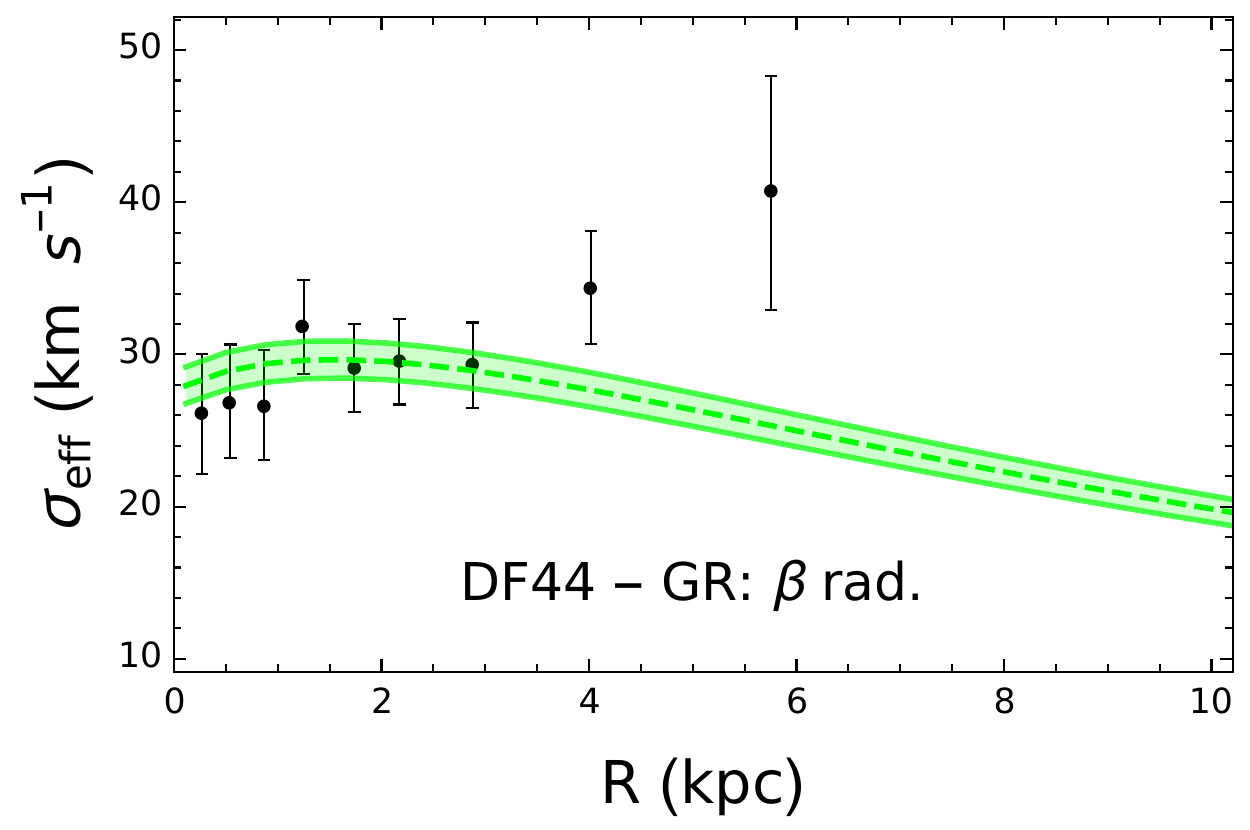}\\
~~~\\
\includegraphics[width=8.cm]{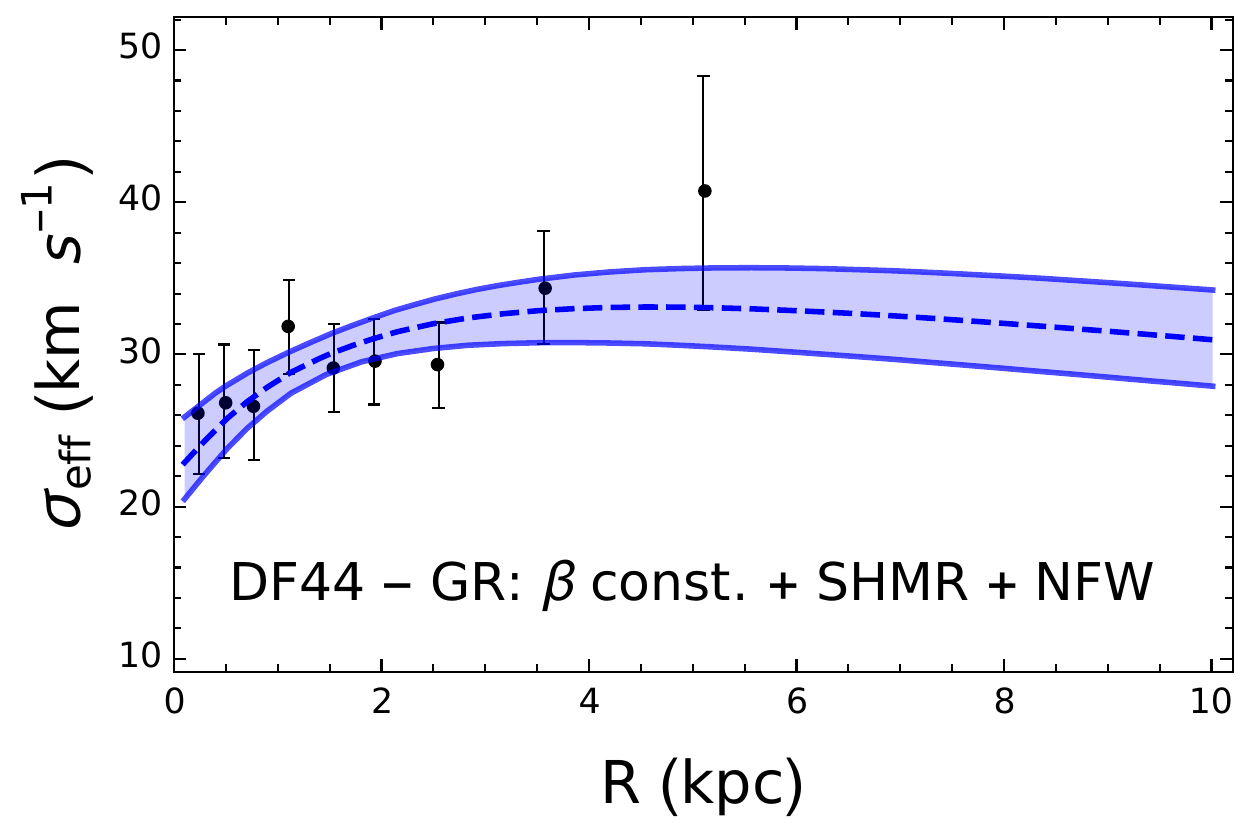}~~~
\includegraphics[width=8.cm]{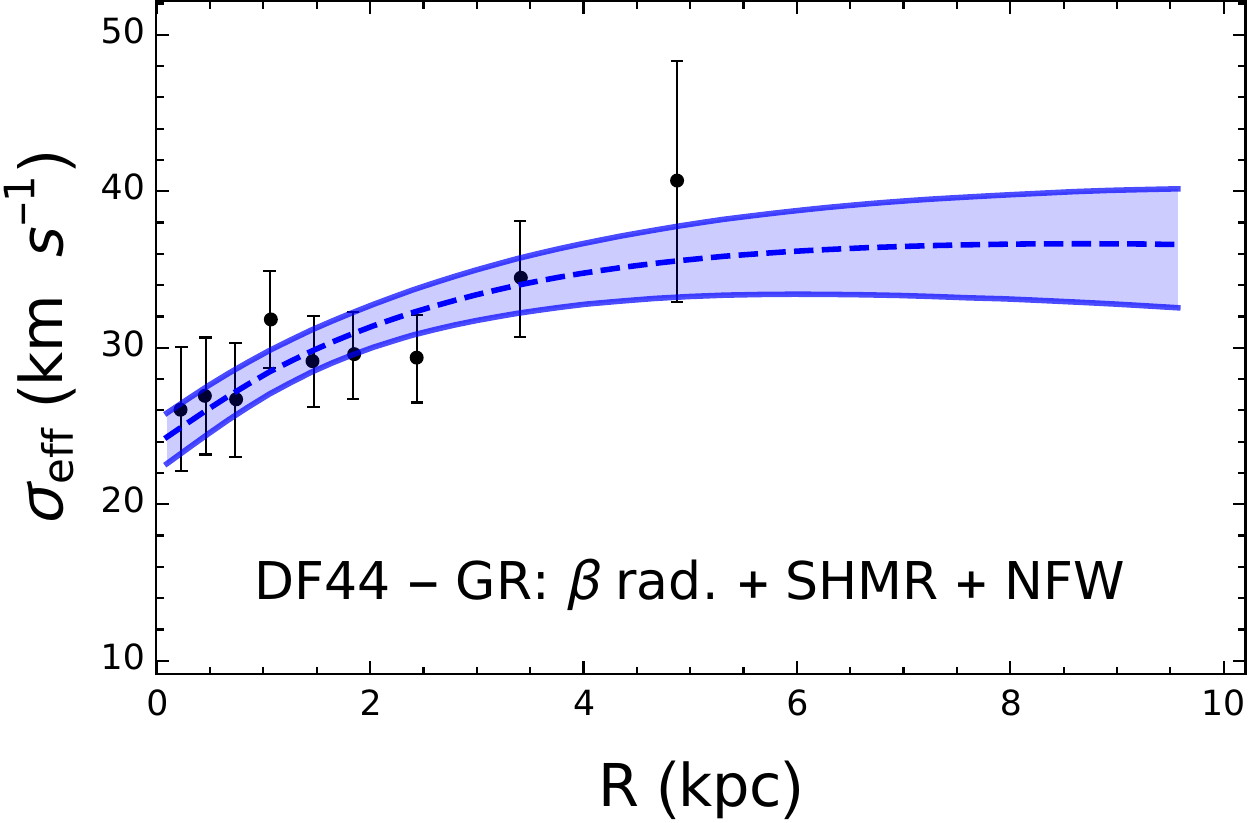}\\
~~~\\
\includegraphics[width=8.cm]{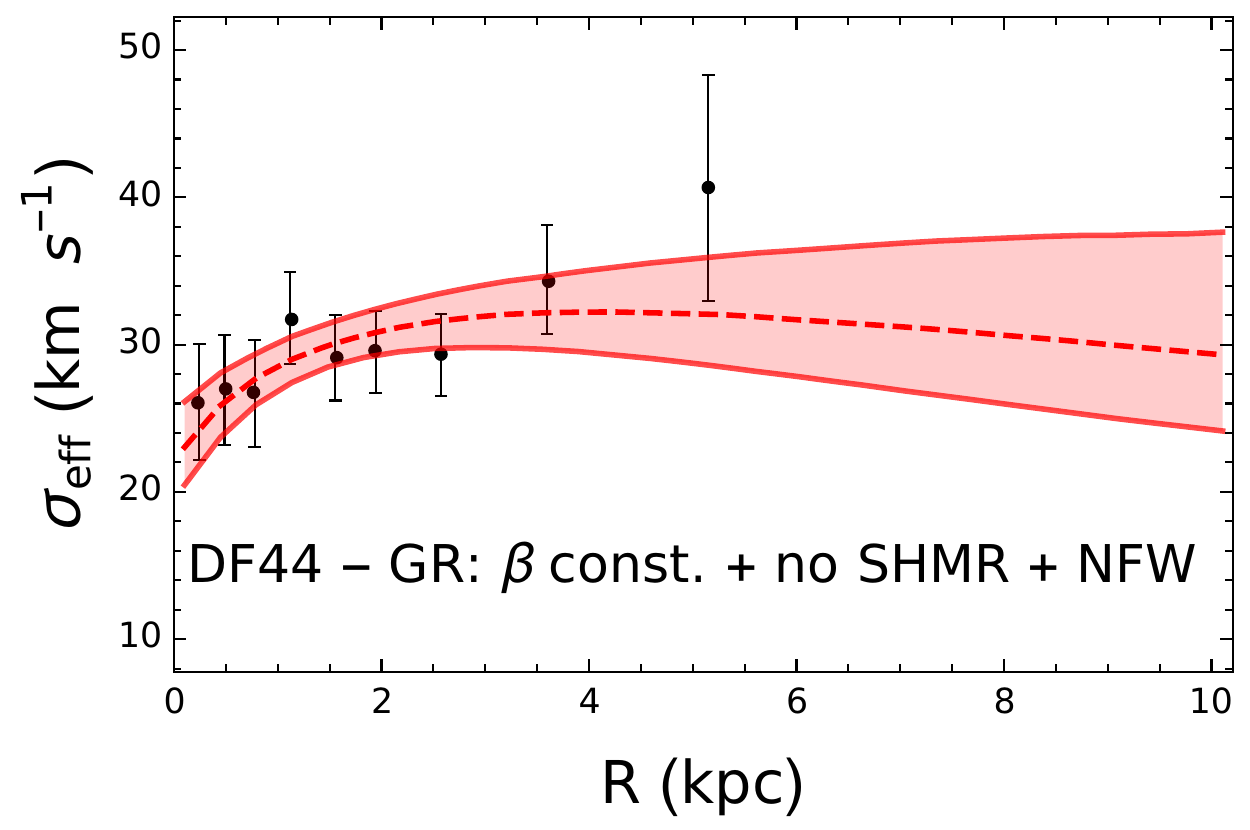}~~~
\includegraphics[width=8.cm]{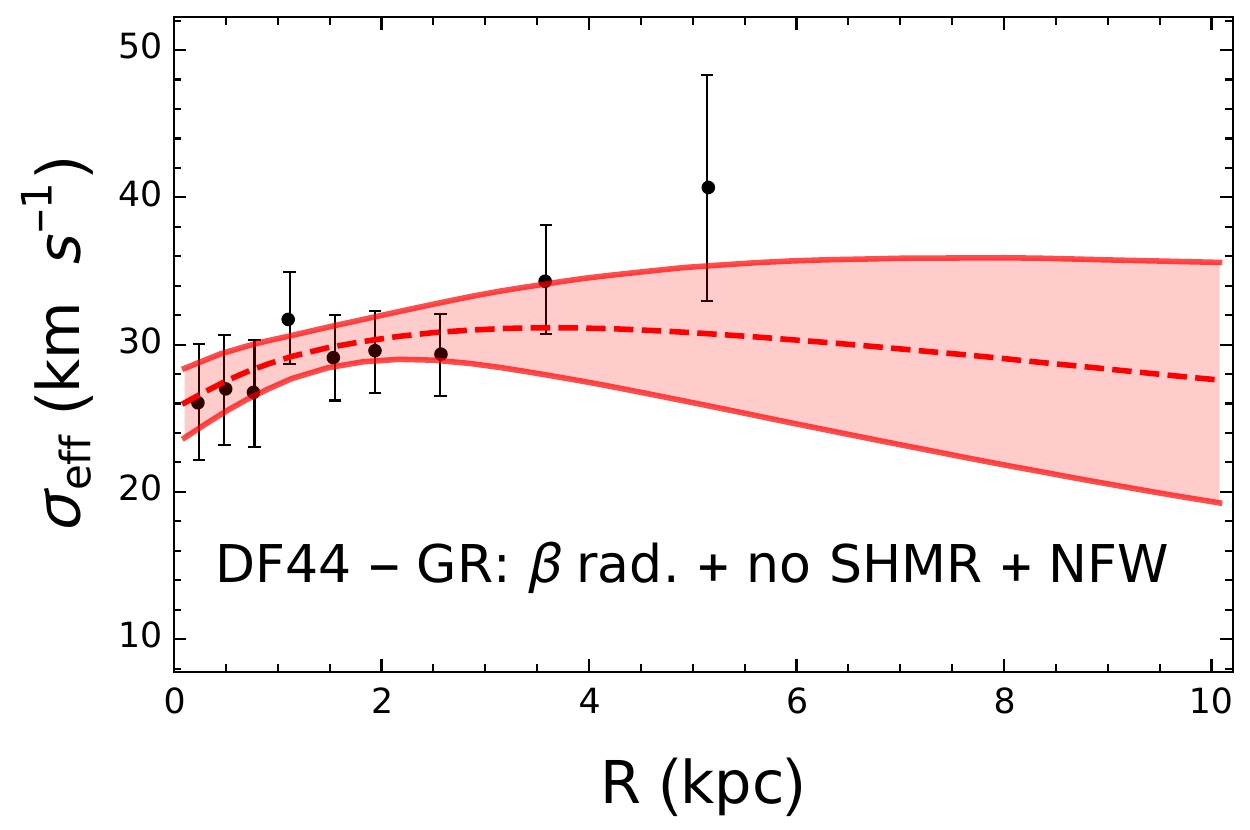}
\caption{Velocity dispersion profiles of Dragonfly 44 with GR. Black dots and bars are observational data, $\sigma_{eff}=\left(\sigma^{2}_{i} + v_{i}^{2}\right)^{1/2}$, with uncertainties. Colored dashed lines and shaded regions are respectively the median and the $1\sigma$ confidence region of the $\sigma_{eff}$ profile derived from Eq.~\eqref{eqn: finvlos}. Left panel: constant velocity  anisotropy profile. Right panel: radial velocity profile.}\label{fig:plot_DF44_GR}
\end{figure*}

\begin{figure*}
\centering
\includegraphics[width=8.cm]{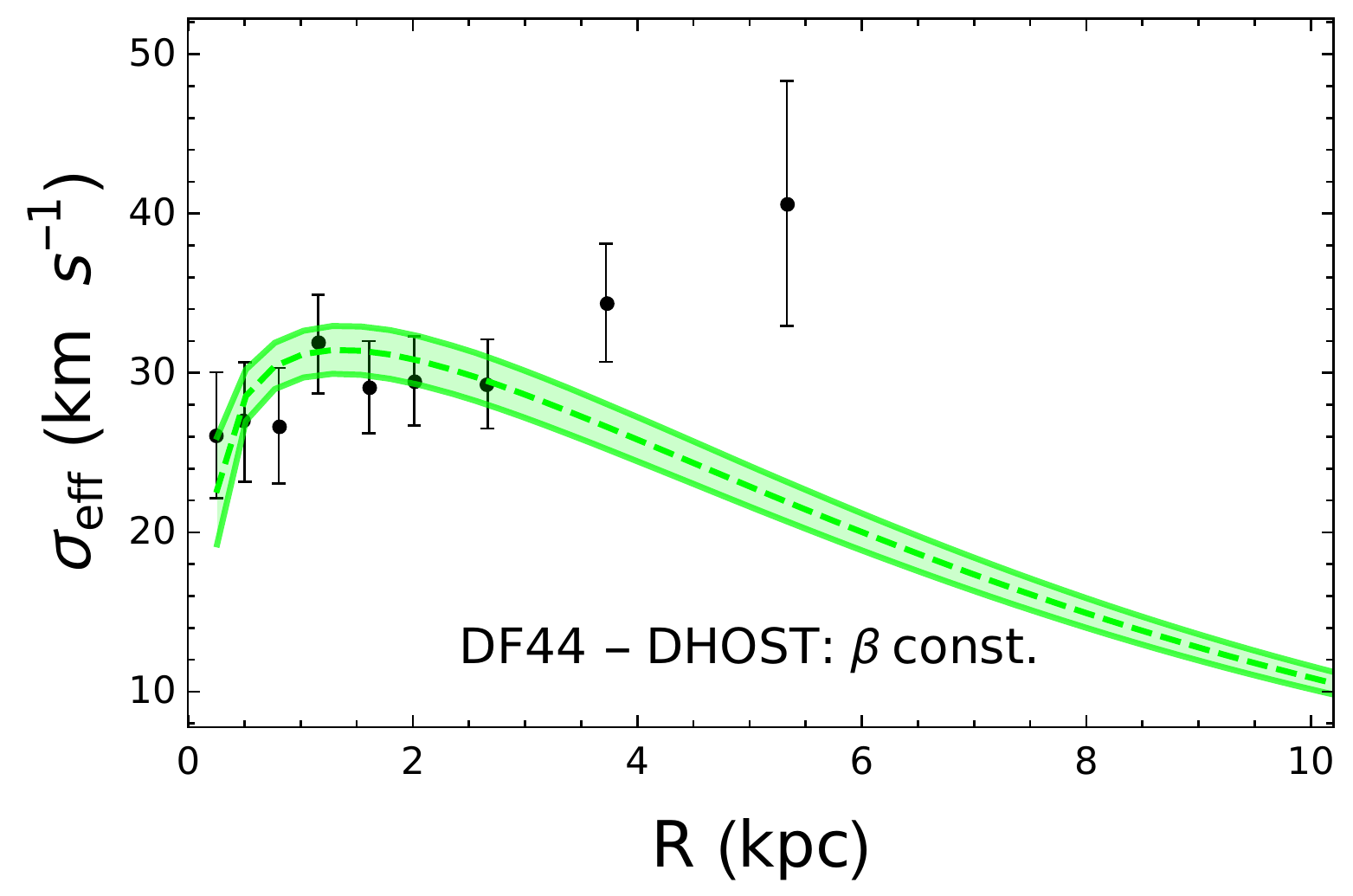}~~~
\includegraphics[width=8.cm]{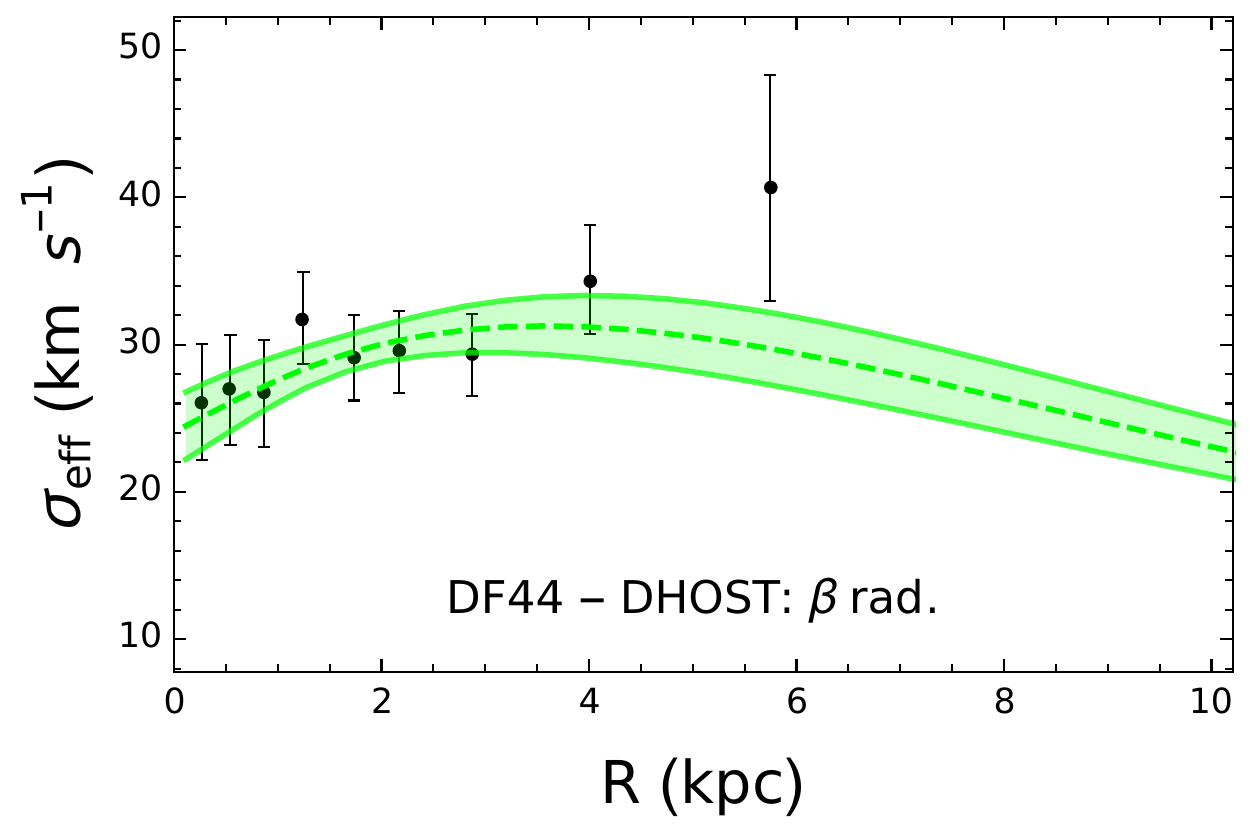}
\\
~~~\\
\includegraphics[width=8.cm]{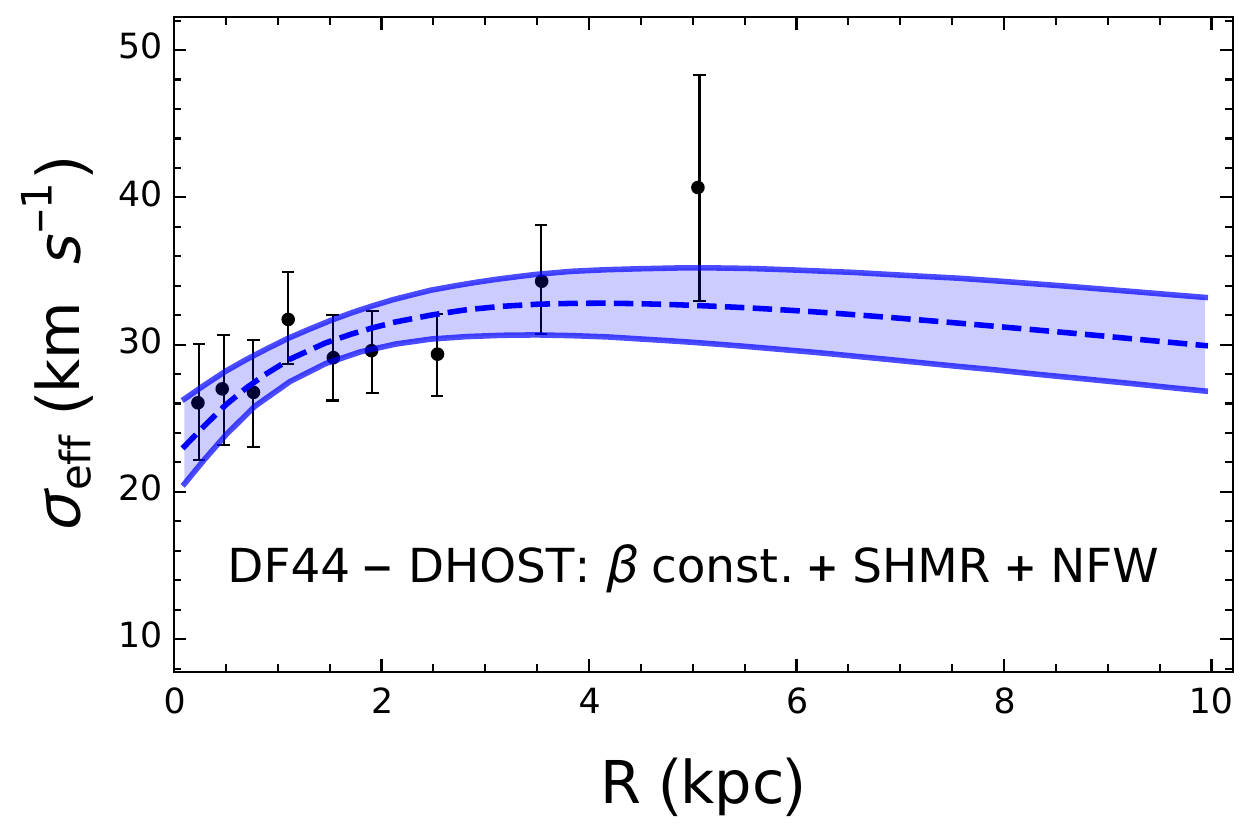}~~~
\includegraphics[width=8.cm]{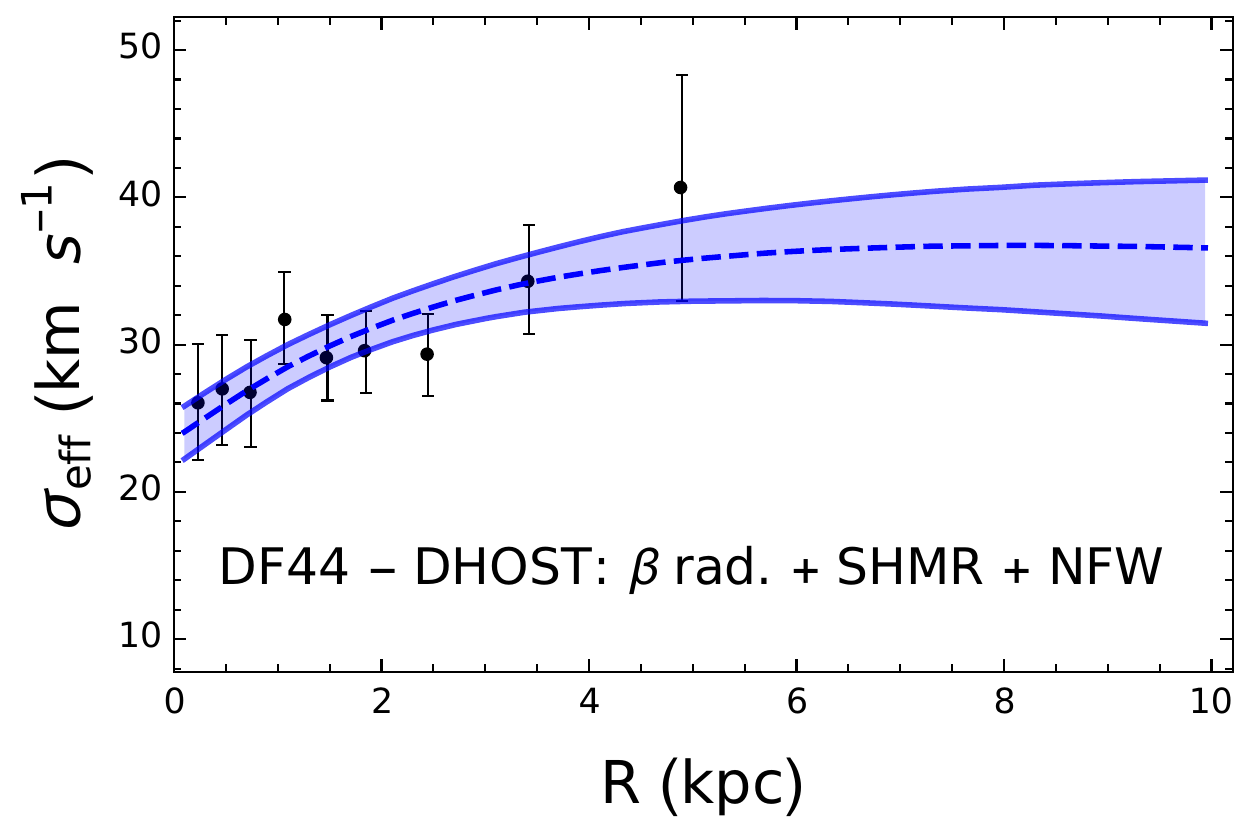}\\
~~~\\
\includegraphics[width=8.cm]{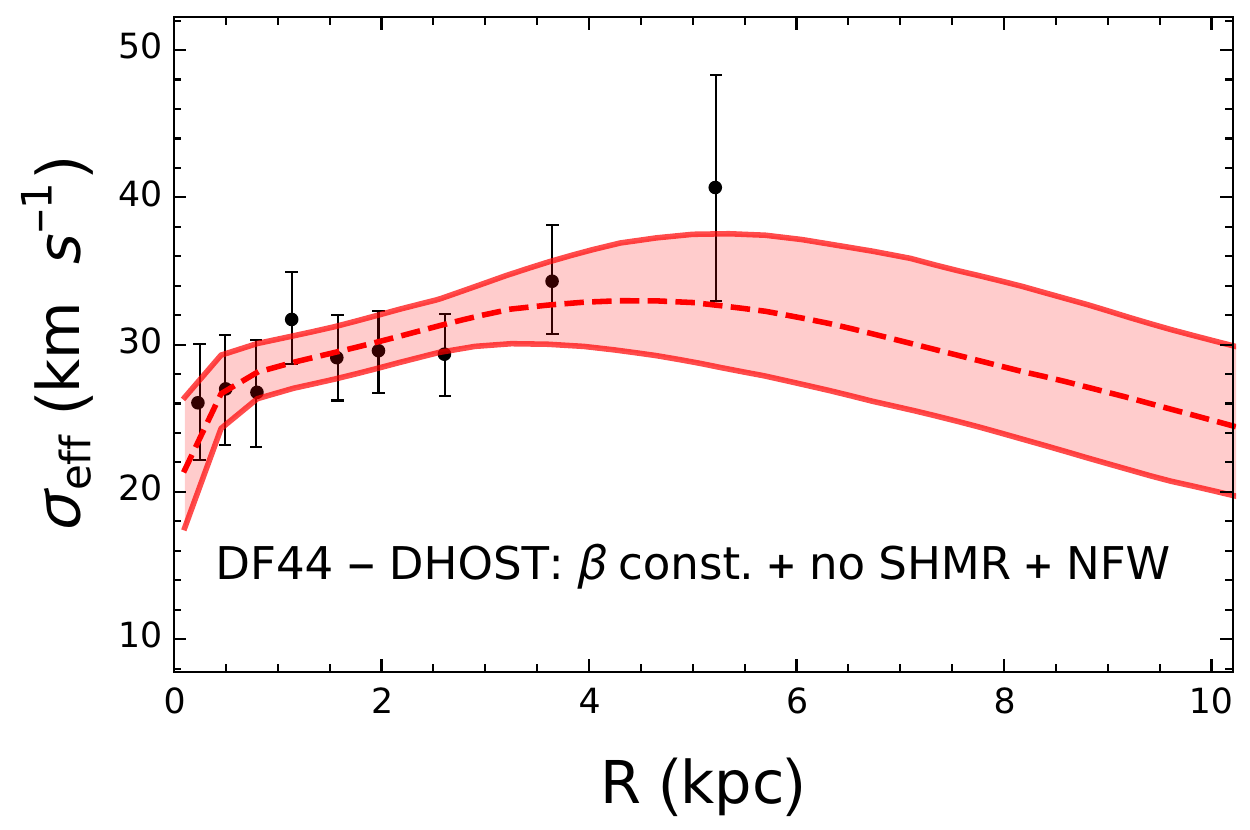}~~~
\includegraphics[width=8.cm]{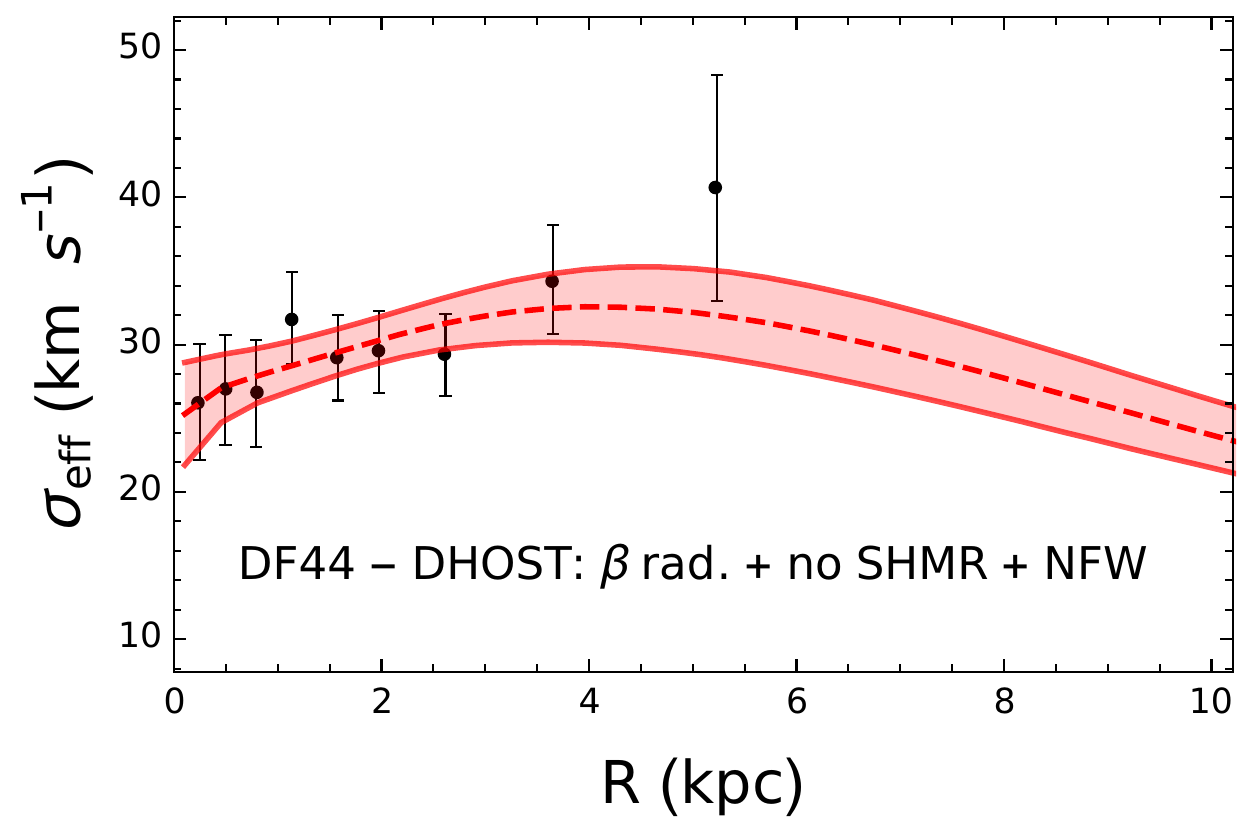}\\
~~~\\
\includegraphics[width=8.cm]{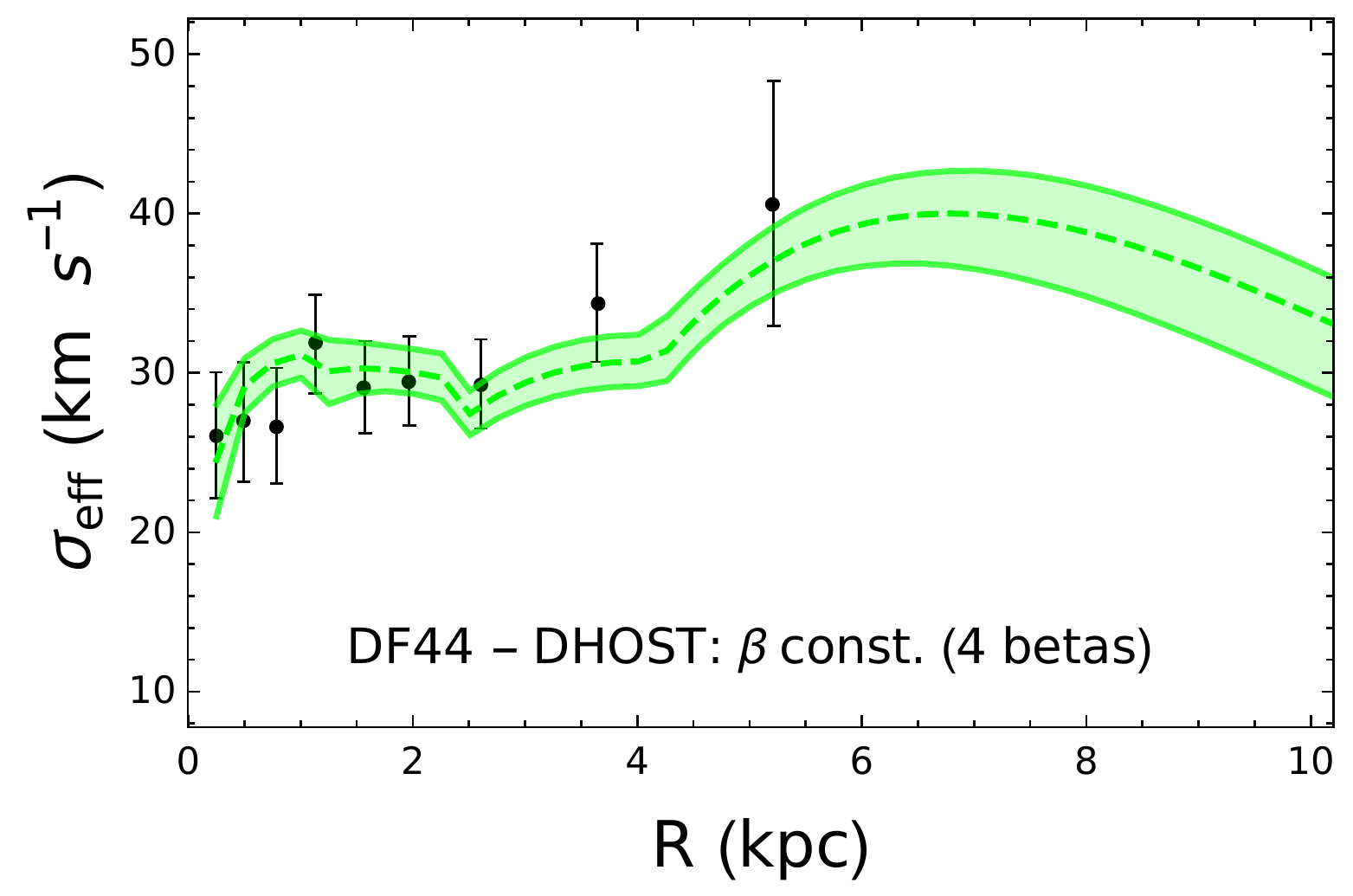}
\caption{Velocity dispersion profiles of Dragonfly 44 with GR. Black dots and bars are observational data, $\sigma_{eff}=\left(\sigma^{2}_{i} + v_{i}^{2}\right)^{1/2}$, with uncertainties. Colored dashed lines and shaded regions are respectively the median and the $1\sigma$ confidence region of the $\sigma_{eff}$ profile derived from Eq.~\eqref{eqn: finvlos_DHOST}. Left panel: constant velocity  anisotropy profile. Right panel: radial velocity profile.}\label{fig:plot_DF44_DHOST}
\end{figure*}

The outputs of the case with GR and only a baryonic component are effectively similar to those found for DF2 \citep{Laudato:2022vmq}, namely, apparently, the kinematics does not require any DM to be supported, as can be seen in the top panel of Fig.~\ref{fig:plot_DF4_GR}. Independently of the functional form, the anisotropy parameters $\beta$ point towards a more tangential anisotropy profile, although the constant anisotropy profile seems to be slightly favored, as can be seen from the values of the Bayes Factors and Suspiciousness (two last columns of Table~\ref{tab:results1}). 

The spatially averaged velocity dispersion is $\sigma\sim 6$ km s$^{-1}$ for the constant anisotropy case and $\sigma\sim 5.9$ km s$^{-1}$  for the case of radial anistropy, both consistent with the estimation of $\sigma \sim 7$ km s$^{-1}$ in \citep{2019ApJ...874L...5V}.

When a DM component is added to the matter budget, we need to distinguish the results according to the presence or not of the SHMR prior. If we look at the middle panels of Fig.~\ref{fig:plot_DF4_GR}, when the prior SHMR is assumed, the velocity dispersion profile is not able to fit data over the entire range of distances for both anisotropy cases, with a spatially averaged value of $\sigma\sim 11-13$ km s$^{-1}$. 

The agreement with the data improves when the prior on the SHMR is relaxed, and the stellar and DM components are decoupled. Indeed, allowing a less amount of DM translates into less tension with the data, as can be seen from the bottom panels of Fig.~\ref{fig:plot_DF4_GR}. In addition, when the anisotropy parameter is constant, the resulting velocity dispersion profile matches the data better, in the inner-most region, than that with the radial $\beta(r)$ given by Eq.(\ref{eqn: anis}).

If we focus more carefully on the gNFW parameter values, we can see how DF4 behaves almost as perfectly as DF2. In the SHMR case, we have: perfectly Gaussian constraints on $c_{200}$ and $M_{200}$; only an upper limit for $\gamma$, which is consistent with zero; and the median value of $c_{200}\sim 8$ does not match exactly the median value we would expect from the theoretical $c-M$ relation from \cite{Correa:2015dva}, thus pointing to some hidden tension. When the SHMR prior is removed the concentration parameter $c_{200_c}\sim$ grows, the virial mass $\log_{10}M_{200}/M_{\odot}\sim 6$ decreases and is only upper-bounded, and $\gamma$ is unconstrained. 

We can reinforce our conclusions by looking at the Bayes Factor and the Suspiciousness: the SHMR scenario is strongly disfavoured, while the absence of the SHMR prior is basically consistent with the reference scenario. Both results strongly point to an absence or large deficiency of DM in the probed ranges.

We now turn our attention to the cases in which the gravitational scenario is described by the DHOST model. 

When the DHOST model describes DE alone, so that we need to include DM in the matter budget, as we can see from Table~\ref{tab:results1}, there is no substantial difference with respect to the corresponding GR cases. The presence of a DE effect from the DHOST model apparently has no relevant role in changing the values of the fitting parameters. It is possible to realize that the results are almost entirely analogous (taking into account the $1\sigma$ confidence levels) to those in GR also by looking at the middle and bottom panels of Fig.~\ref{fig:plot_DF4_DHOST}. 

When we look at the values of DHOST parameter $\Xi_1$, these are perfectly consistent with the GR limit. This somehow does not come unforeseen, as we do not expect DE to play a decisive role at galactic scales.

An important difference between the GR and the DE DHOST case, although, becomes apparent when we look at the last columns of Table~\ref{tab:results1}: we can see that both the Bayesian Factor and the Suspiciousness are improved in the DHOST scenario with respect to the GR framework. In particular, the most significant changes are when the SHMR prior is applied, which now is only slightly disfavoured, and consistent within $1\sigma$ with the reference GR case. Thus, somehow the DHOST effect mitigates or plays some role in the internal kinematics of DF4, alleviating the tension with GR.

Finally, we come to the scenario we are most interested in, i.e. when, as a consequence of the Vainshtein screening, the DHOST model might mimic all the effects that would typically be attributed to the DM. 

The results are enlisted in the bottom part of Table~\ref{tab:results1} and shown in the top panels of Fig~\ref{fig:plot_DF4_DHOST}. Comparing the results with the ones we get within the GR framework, it is possible to see that the DHOST model barely implies any substantial change in the values of the parameters. We can also see that the DHOST parameter $\Xi_1$ is still consistent with the GR limit at the $1\sigma$ confidence level.

From the last columns of Table~\ref{tab:results1}, the constant anisotropy scenario is characterized by a positive Bayesian ratio and a slightly positive Suspiciousness, thus being perfectly equivalent to the GR case in terms of statistical reliability. The case with radial anisotropy instead has a slight improvement which also makes it consistent within $1\sigma$ with the reference case.

\subsection{Dragonfly 44}

For DF44, first of all we consider the case where the dynamical mass of the galaxy is entirely made of baryons. Looking at the first two panels of Fig.~\ref{fig:plot_DF44_GR}, it is clear that the fits with the only baryonic component are very poor. Looking at Table~\ref{tab:results2}, it is possible to notice that the constant anisotropy case implies a higher distance and, in particular, a mass-to-light ratio $\Upsilon_*$ which is not consistent with the range provided in \citep{vanDokkum:2019fdc,Wasserman:2019ttq}. The discrepancy is reduced when assuming a radial anisotropy profile, but still the Bayesian quantities underline the impossibility of the baryonic component, independently from the specific functional form of the anisotropy, to reproduce the observational data.
    

When DM is included in the mass budget, the GR case with a constant anisotropy profile points toward a tangential value for $\beta_c$, which perfectly agrees with the results of \citep{Wasserman:2019ttq}. 
For DM the comparison cannot be so straight because in \citep{Wasserman:2019ttq} only a standard NFW profile has been assumed (corresponding to $\gamma = 1$). If we look at the columns Table~\ref{tab:results2} with the results for the DM parameters, we may notice one first important point: when a constant anisotropy profile is assumed, the reported $c_{200}$ value is fully in agreement with the value which would come from using the $c-M$ relation of \cite{Correa:2015dva}. Thus, there is no tension in this case, differently from what was happening with DF4 and DF2. The upper bound on $\gamma$ is also more relaxed and less cuspy with respect to that described by a standard NFW density profile. 

This scenario is characterized by a rise in the velocity dispersion at larger radii which is in agreement with the data, thus we can unequivocally state that the presence of DM is strongly required. The spatially averaged velocity dispersion we find is $\sigma\sim 31$ km s$^{-1}$, in agreement with the value of $\sigma \sim 33.5$ km s$^{-1}$ presented in \citep{vanDokkum:2019fdc}.

When the anisotropy profile is a radial function, the $\beta$ parameters $\beta_{0,\infty}$ tend also to tangential values and the scale parameter $r_a$ is practically unconstrained, and there seems to be again a small tension between the obtained and the expected values of $c_{200}$. From the top panels of Fig~\ref{fig:plot_DF44_GR}, when we compare the velocity dispersion profiles $\sigma_{eff}$, we can see that the case with $\beta(r)$ exhibits a more significant increase in the velocity dispersion profile for $r \geq 5$ kpc.

When we remove the SHMR prior, allowing for less DM, the situation is not that different from the one described above, but it is totally different from DF4 and DF2. Indeed, looking at Table~\ref{tab:results2}, we can notice how now $c_{200}$ and $M_{200}$ are perfectly consistent with results from literature and with theoretical expectations. The inner log-slope parameter $\gamma$ appears to be substantially unconstrained for the constant anisotropy case, while for the other case it points towards a cored DM halo. From the last columns of Table~\ref{tab:results2}, the Bayes Factor and the Suspiciousness underline full consistency for the constant anisotropy case, which is even slightly favored with respect to the case with applied SHMR. 
The radial anisotropy scenario does not depict a different picture with respect to that in which the SHMR prior is assumed.

We now turn our attention to the DHOST gravitational scenario. Looking first at the case where the DHOST only plays the role of DE, Table~\ref{tab:results2} emphasizes the fact that the presence of DHOST effects does not produce a change in the results we previously got for the GR case, at least in the case in which the SHMR prior is applied. Instead, when this is not included, the values of $c_{200}$ and $M_{200}$ show the same tension which was in DF4 and DF2. Moreover, even the $\gamma$ parameter is now totally different: while so far we always had $\gamma<1$, now it is preferentially $>1$. The values of the DHOST parameter $\Xi_1$ are consistent with zero for the SHMR case, while in the case of no SHMR prior they are quite large, being consistent with zero only at $2\sigma$ level.

When we look at the Bayes Factors and the Suspiciousness, we see how the SHMR cases and the constant anisotropy profiles are the best candidates, with marginal or almost null disfavor with respect to the GR cases.




Finally, when the DHOST model is cast as a DM description, from Table~\ref{tab:results2}, it is possible to notice that the $\beta_c$ case points towards a radial system, and the corresponding value of the $\Xi_1$ parameter is not consistent with the GR limit within the $3\sigma$ confidence level. In the case of a radial anisotropy, we recover a preferentially tangential behavior, but with a slightly large estimation for the distance and the stellar mass-to-light ratio, while the DHOST parameter is now consistent with the GR limit within a $2\sigma$ confidence level.

The top panels of Fig.~\ref{fig:plot_DF44_DHOST} anyway clearly show how in this context the DHOST is unable to replace DM, as well as both the Bayesian Factors and the Suspiciousness point to a strong and significant tension between this scenario and the reference GR case. Definitely, a standard DM component is needed and the DHOST cannot play effectively its role.


Nevertheless, it seems to be important and interesting to highlight here that (when we try to mimic DM with the DHOST) when a radial profile is assumed, there is a rise in the dispersion velocity at larger distances from the center of the galaxy, for which the fit is improved with respect to the corresponding GR case (although still statistically disfavoured with respect to the reference case which includes DM). Given the fact that the anisotropy function $\beta(r)$ is basically unknown, we have tried different functional forms to check if we could find one which adapted better to the data. Here we will describe quickly all the cases, but we will focus more only on what resulted to be the best one, which we show in Table~\ref{tab:results3} and in the bottom panel of Fig.~\ref{fig:plot_DF44_DHOST}.

As first check, we have studied the impact (if any) of the priors on the final results. We have relaxed the Gaussian priors on $\log\left(1-\beta_{0,\infty}\right)$ and we also decided to narrow the flat prior on $r_a \in [0,1]$. However, from a statistical point of view, these new conditions did not produce any changes with the cases already listed in Table~\ref{tab:results2}. We have then studied the feasibility of the Osipkov-Merritt profile \citep{1979SvAL....5...42O,1985MNRAS.214P..25M}
\be
\label{eqn: Osi_Merr}
\beta(r) = \beta_{OM}\frac{r^2}{r^2 + r^2_a}\,.
\ee
Differently from other works \citep{Mamon:2012yb, Mamon:2004xk}, we have left the multiplicative factor $\beta_{OM}$ as a free parameter imposing a flat prior $\beta_{OM}\in [-10,1]$. Nevertheless, the Osipkov-Merritt profile has resulted in a poorer match with the observational data with respect to the cases we already considered.

As ultimate rationale we have explored a ``piecewise'' anisotropy profile, i.e. we have assumed that $\beta(r)$ could be constant or radial within different spatially separated bins. We have explored many different configurations, splitting data from two to four bins, with this last one which resulted to be to the best match, really improving the standard scenarios considered in Table~\ref{tab:results3}. 

While the assumption of a radial profile (in each bin) generally produces worse fits to data, when the anisotropy is assumed to be constant we get interesting results. Within GR, we do not really get any statistically significant change with respect to Table~\ref{tab:results2}, with similar or sometimes slightly worse values for the Bayes Factor and the Suspiciousness. The same consideration holds true when we turn our attention to the DHOST gravitational scenario when it reproduces the DE effects.

The most relevant differences regard the case in which the DHOST model acts as a replacement for DM. Indeed in this case the Suspiciousness moves from a strongly negative value of $\sim -3$ to a slightly positive one of $\sim0.1$. More in detail: the distance and the stellar-to-light mass ratio are perfectly consistent with the literature; the anisotropy profile seems to point to a constant in the first two bins, with a decrease in the outer ones (where we have fewer points and thus only a lower limit can be fixed); and, finally, the characteristic DHOST parameter, $\Xi_1$, is consistent with the result in Table~\ref{tab:results2}. If we look at the bottom panel of Fig.~\ref{fig:plot_DF44_DHOST}, now the DHOST model seems to be able to fit the observational data as successfully as GR.

\begin{figure*}
\centering
\includegraphics[width=17.5cm]{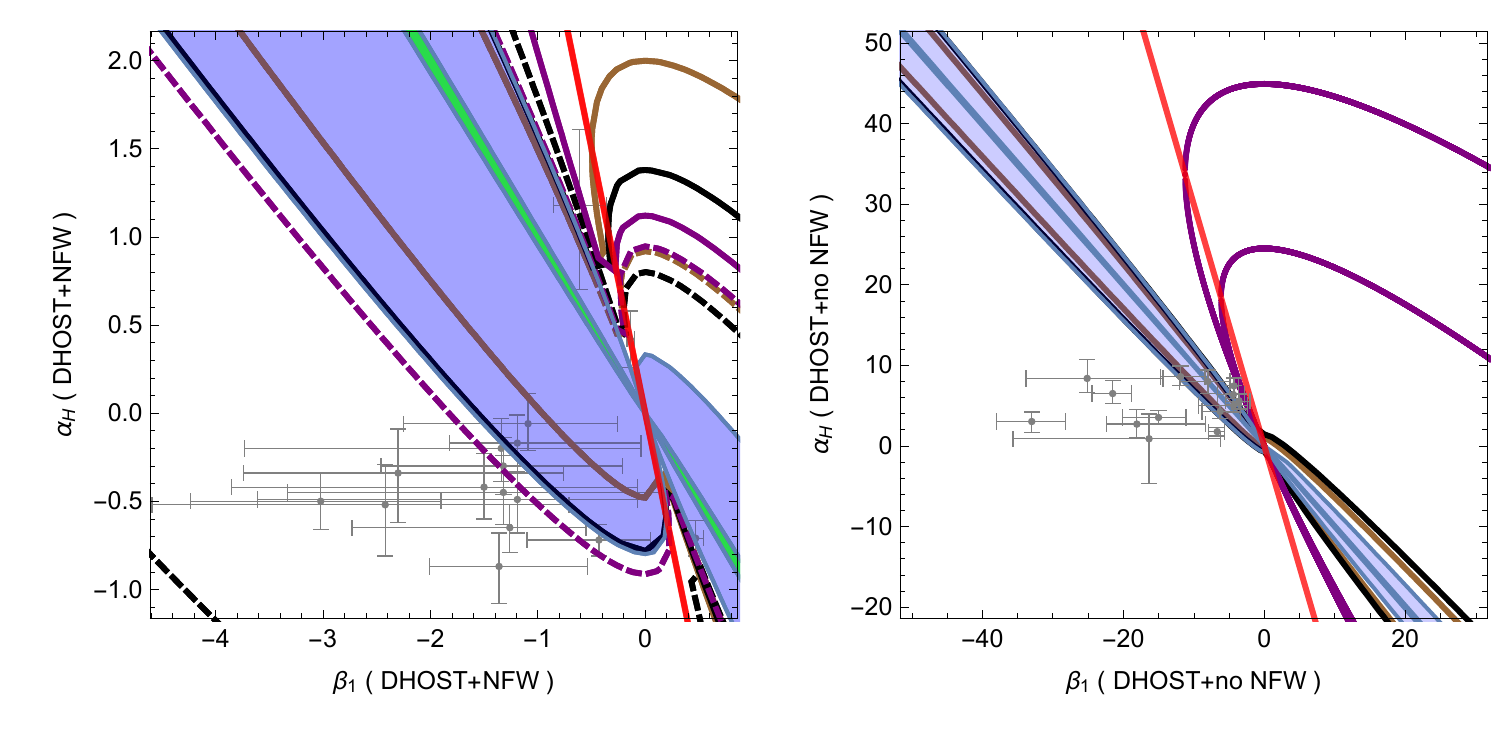}
\caption{Comparison of ETF parameter constraints from our DHOST analysis with the results from \cite{Dima:2017pwp}, \cite{Saltas:2019ius} and \cite{10.1093/mnras/stac180}. Left panel: results under the assumption that DHOST mimicks dark energy. Dashed lines are $1\sigma$ constraints from this work for the case with SHMR prior; solid lines are $1\sigma$ constraints from this work for the case without the SHMR prior. Black is for NGC1052-DF4 and purple is for Dragonfly44. Right panel: results when DHOST is assumed to play the role of both dark energy and dark matter. Solid lines are $1\sigma$ constraints from this work. Black is for NGC1052-DF4 and purple is for Dragonfly44. In all cases we assume a constant stellar anisotropy profile. In both panels blue regions are derived from stellar physics considerations \cite{Dima:2017pwp}; the red region is derived from $2\sigma$ limits on $\gamma_0$ from the Hulse-Taylor pulsar; helioseismology $2\sigma$ constraints \cite{Saltas:2019ius}
are shown as green regions. Single constraints from CLASH clusters as obtained by \cite{10.1093/mnras/stac180} are shown as grey points/crosses. Brown lines are for NGC1052-DF2 from \cite{Laudato:2022vmq}. } \label{fig:EFT}
\end{figure*}

Finally, some remarks concerning the EFT parameters $\alpha_H$ and $\beta_1$, whose constraints can be derived from $\Xi_1$ and compared with the literature \citep{Dima:2017pwp,Saltas:2019ius,10.1093/mnras/stac180}. In particular, in Fig.~\ref{fig:EFT} we report: in blue, the constraints that derive from astrophysical arguments (i.e. dynamical equilibrium \cite{Saito:2015fza} and red dwarf's minimal mass \citep{PhysRevLett.115.201101}); in red, the $2\sigma$ constraints on the parameter $\gamma_0$ from the Hulse-Taylor pulsar \citep{BeltranJimenez:2015sgd,Weisberg:2010zz}; in green, the most stringent constraints on $\alpha_H$ and $\beta_1$ provided by helioseismology arguments \citep{Saltas:2019ius}; in gray, results we achieved in \citep{10.1093/mnras/stac180} from the sample of \textit{CLASH} galaxy clusters; in brown, the constraints recently determined from DF2 in \citep{Laudato:2022vmq}. The new constraints on $\alpha_H$ and $\beta_1$ provided by this work are represented respectively by black lines for DF4 and purple ones for DF44. Dashed lines are when the prior SHMR is assumed, and solid ones when the SHMR is relaxed. 

In the left panel of Fig.~\ref{fig:EFT}, we present the constraints when the DHOST model is assumed to play the role of DE alone. In general, it is possible to see that the constraints from both galaxies span slightly broader ranges than all those covered by stellar constraints, being consistent with them. They also contain both the $\Lambda$CDM limit and the tightest constraint so far from pulsars. 

In the right panel of Fig.~\ref{fig:EFT} we present the constraints on the DHOST parameters $\alpha_H$ and $\beta_1$ when the DHOST model resembles also DM. It is possible to notice that the constraints derived for DF4 are similar to those previously obtained for DF2. This fact is not unexpected since the galaxies are similar in several aspects, such as size, surface brightness, and DM content. But it is also clearly evident that Dragonfly 44 produces constraints that are for a large part not consistent with the other probes which have been considered so far, even in the case of four $\beta$ bins, which still fit the data. This is a further indication of the difficulty of the DHOST model to describe DM in this galaxy.

\section{Conclusions}
\label{sec: conclusions}

In this work we analyze two Ultra-diffuse galaxies, NGC1052-DF4 \citep{2019ApJ...874L...5V} and Dragonfly 44 \citep{vanDokkum:2019fdc} observed with the Dragonfly Telescope Array \citep{vanDokkum:2014cea}. These two galaxies are diametrically opposed for what concerns their estimated DM content: NGC1052-DF4 is characterized by a low-velocity dispersion that underlines a low DM content; Dragonfly 44, instead, appears to be a DM dominated galaxy, since its halo should represent $\sim 99\%$ of the galaxy's total mass. 

We have inferred the mass properties of NGC1052-DF4 and Dragonfly 44 using the Jeans equation Eq.~\eqref{eqn: Jeans} and modeling the galaxies' masses as the sum of a stellar and a DM component (when assumed). Furthermore, we have also considered two different phenomenological models for the anisotropy function $\beta$: one in which it is constant; the other in which it radially changes following Eq.~\eqref{eqn: anis}. 

We have tested two different scenarios within the context of modified gravity theories by choosing a specific model belonging to the family of DHOST theories and characterized by the gravitational potentials as in Eq.~\eqref{eqn: model}. In the first scenario, the modification of the gravity sector is introduced to mimic DE behavior; in the second one, the DHOST effects are extended so to tentatively replace DM at galactic scales as a direct consequence of the partial breaking of the corresponding screening mechanism.

For NGC1052-DF4 we have found results that are basically equivalent to those for NGC1052-DF2 \citep{Laudato:2022vmq}, of which it can be considered a twin: the baryonic galaxy case, with a constant anisotropy, is preferred from a statistical point of view over all other scenarios. The introduction of DM has the net effect of making the Bayesian tools (Bayer Factor and Suspiciousness) worse. This is especially evident for the cases in which a correlation between the stellar and the DM counterpart (the SHMR prior) has been considered. The discrepancy is lighter when the star and DM components are decoupled.

When the DHOST model acts only as DE, it does not produce significant differences compared to the corresponding GR cases. The main effect is an improvement in the Bayesian indicators for all the anisotropy parameter cases, more evident for the SHMR scenario. Generally, we have consistency with GR at $1\sigma$ confidence level. When the DHOST model acts also as a replacement for DM, we find out that the DHOST and GR are still statistically equivalent with basically no tension.

Thus, joining our results from both NGC1052-DF2 and NGC1052-DF4, we may conclude that the DHOST model is apparently successful in describing the dynamics of highly DM-deficient or almost fully-baryonic galaxies. Although, this cannot lead us to state an unequivocal claim in favor of the DHOST model. We can only conclude that, if there is no DM, the DHOST model with only a stellar mass component is consistently able to reproduce the internal dynamics of such galaxies. 

For what concerns galaxies like Dragonfly 44, which are apparently dominated by DM, results are quite different. In general, the best match with the data is provided when GR describes gravity, with a constant anisotropy scenario, and with stellar and DM components being decoupled. 

When the DHOST is taken into account as a DE component, we do not get substantial differences in comparison with the corresponding GR cases. Thus, at least as DE, the DHOST is statistically equivalent to GR. But when the DHOST model tries to replace DM, the Bayesian indicators show a high tension for both the constant ($\beta_c$) and the radial anisotropy described by Eq.~(\ref{eqn: anis}), with respect to the GR reference case. Thus, we can conclude that the DHOST model is not able to replace DM in a system like DF44. However, if the anisotropy parameter is a ``piecewise'' function, allowing for different constant values within different bins, the situation can change, with the DHOST model now seemingly being able to play the role of DM for a dark matter-dominated galaxy.

Unfortunately, when we consider the analysis from the more fundamental perspective of EFT approach and parameters, the situation in the case of Dragonfly 44 is not so easily settled. While NGC1052-DF4 and NGC1052-DF2 analysis result to be perfectly consistent with the constraint which are set by more stringent and precise probes at stellar scales, independently of the presence or less of a DM component, Dragonfly 44 is more troublesome. If we take our DHOST model only as DE, the agreement is quite good; but when we try to replace DM, the constraints on the EFT parameters are totally not consistent with the stellar limits. Thus, this scenario seems to be discarded. 

Further analysis on more systems like Dragonfly 44 should be pursued, to carefully check if and how this inconsistency can be solved or not. Possible candidates for this purpose could be the so-called galaxies which are gas-rich and for this reason called \textit{HI bearing Ultra-diffuse galaxies} \citep{2017ApJ...842..133L}. Within the GR description of gravity, kinematics analysis of some of them has already been realized in \citep{PinaMancera:2021wpc, Shi:2021tyg, Kong:2022oyk}. Thus, an analysis in the context of the DHOST model which we have considered in this work might be interesting.

{\renewcommand{\tabcolsep}{1.5mm}
{\renewcommand{\arraystretch}{2.}
\begin{table*}
\begin{minipage}{\textwidth}
\huge
\centering
\caption{Results from the statistical analysis of NGC1052-DF4. For each parameter we provide the median and the $1\sigma$ constraints; unconstrained parameters are in italic font. The parameters are, from left to right: distance $D$; mass-to-light ratio $\Upsilon$; systemic velocity $v_{sys}$; anisotropy function parameters, depending on the model assumed, constant $(\beta_{c})$ or radial from \cite{Zhang:2015pca} $(\beta_{0},\beta_{\infty},r_{a})$; gNFW concentration $c_{200}$, mass $M_{200}$, and inner log-slope $\gamma$; DHOST characteristic scaling $\Xi_1$; Bayes factor $\mathcal{B}_{i}^{j}$; logarithm of the Bayes factor $\log \mathcal{B}_{i}^{j}$; and the suspiciousness $\log \mathcal{S}_{i}^{j}$.}\label{tab:results1}
\resizebox*{\textwidth}{!}{
\begin{tabular}{c|ccc|cccc|cccc|cc}
\hline
\hline
 & \multicolumn{13}{c}{GR}   \\
\hline
 & $D$ & $\Upsilon_{\ast}$  & $v_{sys}$ & $\beta_c$ & $\beta_0$ & $\beta_{\infty}$ & $r_a$ & $c_{200}$ & $\log M_{200}$ & $\gamma$ & $\Xi_1$ & $\log \mathcal{B}^{i}_{j}$ & $\log \mathcal{S}^i_j$\\
 & Mpc &  & km/s &  &  &  & kpc &  & M$_{\odot}$ &  &  &  &  \\ 
\hline
\multirow{2}{*}{Stars only} & $22.05^{+1.17}_{-1.19}$ & $1.93^{+0.49}_{-0.49}$ & $1445.92^{+2.30}_{-2.26}$ & $>-3.06$ & $-$ & $-$ & $-$ & $-$ & $-$ & $-$ & $-$ & $0$ & $0$\\
& $22.04^{+1.23}_{-1.23}$ & $1.81^{+0.51}_{-0.51}$ & $1445.56^{+2.91}_{-2.99}$ & $-$ & $>-3.37$ & $>-1.54$ & $\mathit{50.47}$ & $-$ & $-$ & $-$ & $-$ & $-0.43^{+0.02}_{-0.02}$ & $-0.10^{+0.05}_{-0.05}$\\
\hline
\multirow{2}{*}{SHMR + gNFW} & $21.90^{+1.17}_{-1.18}$ & $1.91^{+0.49}_{-0.52}$ & $1445.78^{+4.59}_{-4.59}$ & $>-1.98$ & $-$ & $-$ & $-$ & $8.87^{+3.83}_{-2.61}$ & $10.77^{+0.15}_{-0.17}$ & $<0.48$ & $-$ & $-2.25^{+0.02}_{-0.02}$ & $-1.69^{+0.07}_{-0.05}$\\
& $21.83^{+1.24}_{-1.12}$ & $1.89^{+0.52}_{-0.51}$ & $1445.55^{+5.01}_{-5.10}$ & $-$ & $>-1.88$ & $>-1.12$ & $\textit{49.57}$ & $8.53^{+3.69}_{-2.54}$ & $10.78^{+0.16}_{-0.16}$ & $<0.45$ & $-$ & $-2.60^{+0.02}_{-0.02}$ & $-1.83^{+0.07}_{-0.10}$\\
\hline
\multirow{2}{*}{NO SHMR + gNFW} & $22.03^{+1.23}_{-1.15}$ & $1.92^{+0.50}_{-0.49}$ & $1446.04^{+2.39}_{-2.40}$ & $>-3.18$ & $-$ & $-$ & $-$ & $25.63^{+11.61}_{-8.35}$ & $<6.08$ & $\mathit{1.00}$ & $-$ & $-0.21^{+0.01}_{-0.01}$ & $-0.08^{+0.02}_{-0.03}$\\
& $21.99^{+1.20}_{-1.17}$ & $1.83^{+0.52}_{-0.50}$ & $1445.63^{+3.04}_{-3.13}$ & $-$ & $>-3.42$ & $>-1.48$ & $\textit{47.81}$ & $25.72^{+12.83}_{-8.48}$ & $<5.85$ & $\mathit{0.99}$ & $-$ & $-0.65^{+0.01}_{-0.01}$ & $-0.13^{+0.06}_{-0.06}$\\
\hline
\hline
& \multicolumn{13}{c}{DHOST (as dark energy)}   \\
\hline
& $D$ & $\Upsilon_{\ast}$  & $v_{sys}$ & $\beta_c$ & $\beta_0$ & $\beta_{\infty}$ & $r_a$ & $c_{200}$ & $\log M_{200}$ & $\gamma$ & $\Xi_1$ & $\log \mathcal{B}^{i}_{j}$ & $\log \mathcal{S}$\\
 & Mpc &  & km/s &  &  &  & kpc &  & M$_{\odot}$ &  &  &  &  \\ \hline
\multirow{2}{*}{SHMR + gNFW} & $21.98^{+1.21}_{-1.27}$ & $1.91^{+0.50}_{-0.49}$ & $1445.90^{+4.46}_{-4.51}$ & $>-1.70$ & $-$ & $-$ & $-$ & $9.55^{+4.57}_{-3.04}$ & $10.75^{+0.16}_{-0.17}$ & $\textit{0.71}$ & $-0.06^{+1.52}_{-0.34}$  & $-1.72^{+0.03}_{-0.02}$ & $-0.04^{+0.02}_{-0.02}$\\
& $21.84^{+1.24}_{-1.12}$ & $1.93^{+0.53}_{-0.54}$ & $1446.04^{+4.77}_{-4.84}$ & $-$ & $>-2.19$ & $>-1.25$ & $\mathit{47.93}$ & $8.49^{+4.07}_{-2.60}$ & $10.77^{+0.16}_{-0.17}$ & $<0.57$ & $-0.25^{+0.54}_{-0.19}$ & $-2.03^{+0.02}_{-0.02}$ & $-0.43^{+0.16}_{-0.13}$\\
\hline
\multirow{2}{*}{NO SHMR + gNFW} & $22.01^{+1.22}_{-1.21}$ & $1.90^{+0.51}_{-0.49}$ & $1445.90^{+2.35}_{-2.48}$ & $>-2.87$ & $-$ & $-$ & $-$ & $25.68^{+12.15}_{-8.43}$ & $<6.48$ & $\textit{1.03}$ & $-0.09^{+0.48}_{-0.60}$  & $-0.24^{+0.02}_{-0.02}$ & $-0.04^{+0.04}_{-0.04}$\\
& $21.99^{+1.22}_{-1.20}$ & $1.82^{+0.51}_{-0.50}$ & $1445.53^{+2.97}_{-2.98}$  & $-$ & $>-3.34$ & $>-1.46$ & $\mathit{48.05}$ & $25.44^{+11.92}_{-8.36}$ & $<6.23$ & $\mathit{0.99}$ & $-0.22^{+0.62}_{-0.70}$ & $-0.58^{+0.02}_{-0.02}$ & $-0.09^{+0.06}_{-0.06}$\\
\hline
\hline
& \multicolumn{13}{c}{DHOST (as dark matter)}   \\
\hline
& $D$ & $\Upsilon_{\ast}$  & $v_{sys}$ & $\beta_c$ & $\beta_0$ & $\beta_{\infty}$ & $r_a$ & $c_{200}$ & $\log M_{200}$ & $\gamma$ & $\Xi_1$ & $\log \mathcal{B}^{i}_{j}$ & $\log \mathcal{S}$\\
 & Mpc &  & km/s &  &  &  & kpc &  & M$_{\odot}$ &  &  &  &  \\ 
\hline
\multirow{2}{*}{Stars only} & $22.09^{+1.20}_{-1.22}$ & $1.92^{+0.49}_{-0.49}$ & $1445.89^{+2.29}_{-2.31}$ & $>-2.22$ & $-$ & $-$ & $-$ & $-$ & $-$ & $-$ & $-0.11^{+0.48}_{-0.62}$ & $0.03^{+0.02}_{-0.01}$ & $0.04^{-0.04}_{-0.04}$\\
& $21.95^{+1.55}_{-2.63}$ & $1.82^{+0.52}_{-0.51}$ & $1445.69^{+2.91}_{-2.89}$ & $-$ & $>-3.24$ & $>-1.36$ & $\mathit{47.68}$ & $-$ & $-$ & $-$ & $-0.18^{+0.53}_{-0.61}$ & $-0.37^{+0.01}_{-0.01}$ & $-0.02^{+0.06}_{-0.04}$\\
\hline
\end{tabular}}
\end{minipage}
\end{table*}}}

{\renewcommand{\tabcolsep}{1.5mm}
{\renewcommand{\arraystretch}{2.}
\begin{table*}
\begin{minipage}{\textwidth}
\huge
\centering
\caption{Results from the statistical analysis of DF44. For each parameter we provide the median and the $1\sigma$ constraints; unconstrained parameters are in italic font. The parameters are, from left to right: distance $D$; mass-to-light ratio $\Upsilon$; anisotropy function parameters, depending on the model assumed, constant $(\beta_{c})$ or radial from \cite{Zhang:2015pca} $(\beta_{0},\beta_{\infty},r_{a})$; gNFW concentration $c_{200}$, mass $M_{200}$, and inner log-slope $\gamma$; DHOST characteristic scaling $\Xi_1$; logarithm of the Bayes factor $\log \mathcal{B}_{i}^{j}$; the suspiciousness $\log\mathcal{S}_{i}^{j}$.}\label{tab:results2}
\resizebox*{\textwidth}{!}{
\begin{tabular}{c|cc|cccc|cccc|cc}
\hline
\hline
 & \multicolumn{11}{c}{GR}   \\
\hline
 & $D$ & $\Upsilon_{\ast}$  & $\beta_c$ & $\beta_0$ & $\beta_{\infty}$ & $r_a$ & $c_{200}$ & $\log M_{200}$ & $\gamma$ & $\Xi_1$ & $\log \mathcal{B}^{i}_{j}$ & $\log \mathcal{S}^{i}_{j}$\\
 & Mpc &  &  &  &  & kpc &  & M$_{\odot}$ &  &  &  &\\ 
\hline
\multirow{2}{*}{Stars only} & $153.98^{+10.98}_{-10.79}$ & $9.10^{+0.59}_{-0.79}$ & $0.17^{+0.11}_{-0.13}$ & $-$ & $-$ & $-$ & $-$ & $-$ & $-$ & $-$ & $-21.45^{+0.01}_{-0.01}$ & $-21.75^{+0.03}_{-0.02}$\\
& $111.99^{+12.61}_{-12.13}$ & $2.15^{+0.41}_{-0.32}$ & $-$ & $-0.43^{+1.40}_{-1.03}$ & $-0.27^{+0.69}_{-1.06}$ & $\mathit{56.62}$ & $-
$ & $-$ & $-$ & $-$ & $-2.06^{+0.01}_{-0.01}$ & $-2.23^{+0.04}_{-0.04}$\\
\hline
\multirow{2}{*}{SHMR+gNFW} &$99.28^{+13.23}_{-13.59}$ & $1.55^{+0.39}_{-0.32}$ & $-0.40^{+0.28}_{-0.45}$ & $-$ & $-$ & $-$ & $10.71^{+3.21}_{-2.84}$ & $10.97^{+0.15}_{-0.16}$ & $<0.73$ & $-$ & $\textit{0}$ & $\textit{0}$\\
&$94.90^{+14.34}_{-13.53}$ & $1.52^{+0.39}_{-0.32}$ & $-$ & $>-3.06$ & $> -1.77$ & $\mathit{50.41}$ & $8.14^{+2.61}_{-2.08}$ & $10.93^{+0.15}_{-0.16}$ & $< 0.45$ & $-$ &  $-0.24^{+0.02}_{-0.01}$ & $-0.07^{+0.05}_{-0.07}$\\
\hline
\multirow{2}{*}{NO SHMR+gNFW} &$100.23^{+14.69}_{-13.72}$ & $1.62^{+0.42}_{-0.34}$ & $>-1.29$ & $-$ & $-$ & $-$ & $11.89^{+6.94}_{-4.73}$ & $10.55^{+1.36}_{-0.48}$ & $\textit{0.78}$ & $-$ & $0.08^{+0.02}_{-0.02}$ & $0.25^{+0.06}_{-0.05}$\\
& $99.93^{+14.32}_{-15.32}$ & $1.62^{+0.42}_{-0.35}$ & $-$ & $>-4.18$ & $>-1.97$ & $\mathit{48.96}$ & $10.86^{+5.98}_{-3.94}$ & $10.22^{+0.76}_{-0.66}$ & $<0.60$ & $-$ & $-0.22^{+0.01}_{-0.02}$ & $-0.04^{+0.05}_{-0.05}$\\
\hline
\hline
& \multicolumn{11}{c}{DHOST (as dark energy)}   \\
\hline
 & $D$ & $\Upsilon_{\ast}$  & $\beta_c$ & $\beta_0$ & $\beta_{\infty}$ & $r_a$ & $c_{200}$ & $\log M_{200}$ & $\gamma$ & $\Xi_1$ & $\log \mathcal{B}^{i}_{j}$ & $\log \mathcal{S}^{i}_{j}$\\
 & Mpc &  &  &  &  & kpc &  & M$_{\odot}$ &  &  &  &\\ 
\hline
\multirow{2}{*}{SHMR+gNFW} & $98.57^{+13.32}_{-14.08}$ & $1.52^{+0.38}_{-0.31}$ & $-0.32^{+0.38}_{-0.46}$ & $-$ & $-$ & $-$ & $9.98^{+3.52}_{-2.98}$ & $10.95^{+0.15}_{-0.16}$ & $<0.76$ & $-0.019^{+0.515}_{-0.409}$ & $-0.04^{+0.01}_{-0.02}$ & $0.02^{+0.05}_{-0.05}$\\
&$95.31^{+14.51}_{-14.78}$ & $1.52^{+0.38}_{-0.31}$ & $-$ & $>-2.83$ & $>-1.36$ & $\mathit{49.55}$ & $7.87^{+2.81}_{-2.12}$ & $10.95^{+0.14}_{-0.17}$ & $<0.52$ & $-0.08^{+0.30}_{-0.18}$ & $-0.29^{+0.02}_{-0.01}$ & $-0.03^{+0.07}_{-0.06}$\\
\hline
\multirow{2}{*}{NO SHMR+gNFW} & $101.62^{+14.70}_{-14.28}$ & $1.51^{+0.41}_{-0.31}$ & $>-2.70$ & $-$ & $-$ & $-$ & $18.85^{+8.87}_{-5.59}$ & $9.08^{+0.78}_{-0.39}$ & $>1.53$ & $-7.68^{+6.75}_{-6.27}$ & $-0.07^{+0.02}_{-0.02}$ & $0.09^{+0.05}_{-0.05}$\\
& $101.83^{+13.59}_{-13.10}$ & $1.57^{+0.40}_{-0.32}$ & $-$ & $>-6.81$ & $>-2.52$ & $\mathit{50.28}$ & $18.56^{+9.57}_{-6.52}$ & $8.60^{+0.62}_{-0.51}$ & $1.10^{+0.48}_{-0.61}$ & $-1.67^{+1.08}_{-2.11}$ & $-0.49^{+0.01}_{-0.02}$ & $-0.15^{+0.06}_{-0.05}$\\
\hline
\hline
& \multicolumn{11}{c}{DHOST (as dark matter)}   \\
\hline
  & $D$ & $\Upsilon_{\ast}$  & $\beta_c$ & $\beta_0$ & $\beta_{\infty}$ & $r_a$ & $c_{200}$ & $\log M_{200}$ & $\gamma$ & $\Xi_1$ & $\log \mathcal{B}^{i}_{j}$ & $\log \mathcal{S}^{i}_{j}$\\
 & Mpc &  &  &  &  & kpc &  & M$_{\odot}$ &  &  &  &\\  
\hline
\multirow{2}{*}{Stars only} & $103.74^{+14.11}_{-13.68}$ & $1.67^{+0.45}_{-0.36}$ & $0.92^{+0.02}_{-0.02}$ & $-$ & $-$ & $-$ & $-$ & $-$ & $-$ & $-17.10^{+4.66}_{-6.24}$ & $-2.90^{+0.01}_{-0.01}$ & $-3.09^{+0.03}_{-0.03}$\\
& $111.77^{+12.63}_{-12.21}$ & $2.09^{+0.38}_{-0.32}$ & $-$ & $-8.49^{+1.41}_{-1.01}$ & $-0.11^{+0.58}_{-0.95}$ & $\mathit{44.54}$ & $-$ & $-$ & $-$ & $-0.32^{+0.16}_{-0.18}$ & $-1.18^{+0.02}_{-0.02}$ & $-1.27^{+0.03}_{-0.04}$\\
\hline
\end{tabular}}
\end{minipage}
\end{table*}}}

{\renewcommand{\tabcolsep}{1.5mm}
{\renewcommand{\arraystretch}{2.}
\begin{table*}
\begin{minipage}{\textwidth}
\huge
\centering
\caption{Results from the statistical analysis of NGC1052-DF44 with a ``piecewise'' anisotropy profile. For each parameter we provide the median and the $1\sigma$ constraints; unconstrained parameters are in italic font. The parameters are, from left to right: distance $D$; mass-to-light ratio $\Upsilon$; anisotropy function parameters with a division in four bins; gNFW concentration $c_{200}$, mass $M_{200}$, and inner log-slope $\gamma$; DHOST characteristic scaling $\Xi_1$; logarithm of the Bayes factor $\log \mathcal{B}_{i}^{j}$; the suspiciousness $\log\mathcal{S}_{i}^{j}$.}\label{tab:results3}
\resizebox*{\textwidth}{!}{
\begin{tabular}{c|cc|cccc|cccc|cc}
\hline
\hline
& \multicolumn{11}{c}{GR}\\
\hline
& $D$ & $\Upsilon_{\ast}$ & $\beta_1$ & $\beta_2$ & $\beta_3$ & $\beta_4$ & $c_{200}$ & $\log M_{200}$ & $\gamma$ & $\Xi_1$ & $\log \mathcal{B}^{i}_{j}$ & $\log \mathcal{S}$\\
& Mpc &  &  &  &  &  &  & $M_{\odot}$ &  & \\ 
\hline
Stars only & $152.03^{+10.96}_{-10.94}$ & $8.95^{+0.65}_{-0.81}$ & $0.14^{+0.13}_{-0.16}$ & $0.57^{+0.16}_{-0.20}$ & $>-5.81$ & $>$ & $-$ & 
$-$ & $-$ & $-$ & $-19.72^{+0.01}_{-0.01}$ & $-19.90^{+0.04}_{-0.05}$\\
\hline
SHMR + gNFW & $98.31^{+14.11}_{-13.90}$ & $1.53^{+0.39}_{-0.33}$ & $>-1.15$ & $>-2.84$ & $\mathit{-4.45}$ & $\mathit{-4.71}$ & $10.36^{+3.89}_{-3.11}$ & $10.93^{+0.15}_{-0.18}$ & $<1.00$ & $-$ & $-0.39^{+0.02}_{-0.02}$ & $-0.004^{+0.072}_{-0.067}$\\
\hline
no SHMR + gNFW & $102.18^{+13.73}_{-14.48}$ & $1.60^{+0.42}_{-0.34}$ & $\mathit{-4.04}$ & $\mathit{-5.25}$ & $\mathit{-3.70}$ & $\mathit{-5.85}$ & $13.93^{+6.84}_{-4.31}$ & $10.10^{+0.26}_{-0.18}$ & $>1.46$ & $-$ & $-0.20^{+0.01}_{-0.02}$ & $-0.20^{+0.05}_{-0.04}$\\
\hline
\hline
& \multicolumn{11}{c}{DHOST (as dark energy)}\\
\hline
& $D$ & $\Upsilon_{\ast}$ & $\beta_1$ & $\beta_2$ & $\beta_3$ & $\beta_4$ & $c_{200}$ & $\log M_{200}$ & $\gamma$ & $\Xi_1$ & $\log \mathcal{B}^{i}_{j}$ & $\log \mathcal{S}$\\
& Mpc &  &  &  &  &  &  & $M_{\odot}$ &  & \\ 
\hline
SHMR + gNFW &$ 95.27^{+13.32}_{-14.08}$ & $1.48^{+0.38}_{-0.31}$ & $-0.96^{+0.60}_{-0.92}$ & $>-3.87$ & $\mathit{-4.58}$ & $\mathit{-5.04}$ & $9.40^{+3.64}_{-2.67}$ & $10.97^{+0.17}_{-0.18}$ & $<0.97$ & $-0.019^{+0.515}_{-0.409}$ & $-0.04^{+0.01}_{-0.02}$ & $0.02^{+0.05}_{-0.05}$\\
\hline
no SHMR + gNFW &$101.40^{+14.12}_{-12.88}$ & $1.57^{+0.41}_{-0.31}$ & $\mathit{-3.98}$ & $\mathit{-5.83}$ & $\mathit{-4.83}$ & $\mathit{-7.52}$ & $18.28^{+9.08}_{-6.99}$ & $9.64^{+0.59}_{-0.61}$ & $>1.88$ & $-1.88^{+2.19}_{-4.69}$ & $-0.03^{+0.01}_{-0.02}$ & $-0.005^{+0.005}_{-0.005}$\\
\hline
\hline
& \multicolumn{11}{c}{DHOST (as dark matter)}\\
\hline
& $D$ & $\Upsilon_{\ast}$ & $\beta_1$ & $\beta_2$ & $\beta_3$ & $\beta_4$ & $c_{200}$ & $\log M_{200}$ & $\gamma$ & $\Xi_1$ & $\log \mathcal{B}^{i}_{j}$ & $\log \mathcal{S}$\\
& Mpc &  &  &  &  &  &  & $M_{\odot}$ &  & \\ 
\hline
Stars only & $101.33^{+13.98}_{-14.48}$ & $1.54^{+0.42}_{-0.31}$ & $0.93^{+0.02}_{-0.02}$ & $0.90^{+0.06}_{-0.07}$ & $>-0.19$ & $\mathit{-4.89}$ & $-$ & $-$ & $-$ & $-18.50^{+4.93}_{-6.75}$ & $0.07^{+0.02}_{-0.02}$ & $0.10^{+0.05}_{-0.05}$\\
\hline
\end{tabular}}
\end{minipage}
\end{table*}}}

\bibliographystyle{apsrev4-1}
\bibliography{biblio}

\end{document}